\documentclass[conference]{IEEEtran}

\usepackage{xspace}
\usepackage{listings}
\usepackage{multirow}
\usepackage{graphicx}
\usepackage{url}
\usepackage{xcolor}
\usepackage{color}
\usepackage{caption}
\usepackage{subcaption}
\usepackage{enumitem}
\usepackage{soul}
\usepackage{amsmath}
\usepackage[multiple]{footmisc}
\usepackage{booktabs}
\usepackage{hyperref}

\usepackage{listings}
\usepackage{pdfpages}
\usepackage{float}
\usepackage{tcolorbox}
\newtcolorbox{answerbox}{
  colback=white,   
  colframe=black,  
  sharp corners,   
  boxrule=0.8pt,   
  left=6pt,        
  right=6pt,       
  top=4pt,         
  bottom=4pt,      
  before skip=10pt,
  after skip=10pt, 
  fontupper=\normalsize\flushleft 
}
\usepackage[ruled, lined, linesnumbered, commentsnumbered, longend]{algorithm2e}
\usepackage{setspace}
\usepackage{amsmath,amsthm,amssymb}
\newcommand{\hang}[1]{}

\newcommand{\codename}{\mbox{\textsc{SemPRE}}\xspace}

\newcommand{\ignore}[1]{}

\begin{document}

\RestyleAlgo{ruled}
\SetKwComment{Comment}{/* }{ */}

\date{\today()}

\title{\Large \bf \codename: Semantic-Aware Protocol Reverse Engineering via Consistency-Driven Joint Optimization}

\author{\IEEEauthorblockN{Yihua Xu}
	\IEEEauthorblockA{Illinois Institute of Technology\\
		yxu131@hawk.illinoistech.edu}
	\and
	\IEEEauthorblockN{Weijie Huang}
	\IEEEauthorblockA{Indiana University \\
		wh25@iu.edu}
	\and
    \IEEEauthorblockN{Xiaofeng Wang}
    \IEEEauthorblockA{Nanyang Technological University \\
        xiaofeng.wang@ntu.edu.sg }
    \and
    \IEEEauthorblockN{Hang Zhang}
    \IEEEauthorblockA{Indiana University \\
         hz64@iu.edu }
    \and
    \IEEEauthorblockN{Binghui Wang}
    \IEEEauthorblockA{Illinois Institute of Technology\\
		bwang70@illinoistech.edu }
    \and
	\IEEEauthorblockN{Yue Duan}
	\IEEEauthorblockA{Singapore Management University\\
		yueduan@smu.edu.sg }}


\IEEEoverridecommandlockouts
\makeatletter\def\@IEEEpubidpullup{6.5\baselineskip}\makeatother
\IEEEpubid{\parbox{\columnwidth}{
		Network and Distributed System Security (NDSS) Symposium 2027\\
		22--26 March 2027, Seoul, Republic of Korea\\
		ISBN 978-1-970672-09-1\\  
		https://dx.doi.org/10.14722/ndss.2027.242825\\
		www.ndss-symposium.org
}
\hspace{\columnsep}\makebox[\columnwidth]{}}

\maketitle

\begin{abstract}

Protocol reverse engineering (PRE) is essential for understanding undocumented or proprietary communication formats, enabling security-critical tasks such as vulnerability discovery, malware C2 analysis, and IDS signature generation. However, existing PRE techniques rely on brittle hand-written rules, under-model contextual dependencies among neighboring fields, and often fail on messages with nested sub-protocols. We present \codename, a fully automated PRE framework that replaces ad-hoc heuristics with recursive LLM reasoning grounded in fine-grained dynamic program analysis. \codename uses taint-flow and loop traces to guide an LLM-driven tree-of-thought search that slices messages into fields while detecting nested sub-payloads, producing a hierarchical layout. It then infers each field’s type and role and iteratively refines boundaries by freezing high-confidence segments and re-running detection under semantic constraints, yielding an EM-style procedure that monotonically improves structure and semantics until convergence. Evaluated on a broad corpus of mainstream protocols against six state-of-the-art PRE systems, \codename achieves an average F1 of 0.76 for field detection (vs.\ $<$0.40 for all baselines) and 0.72 for semantic inference (vs.\ 0.14 for BinPRE). Ablations confirm the contribution of each component, and a case study on a prolific real-world botnet demonstrates practical utility. 
\end{abstract} 
\section{Introduction}
\label{sec:intro}

\begin{figure}[h]
    \centering
    \vspace{-3mm}
    \includegraphics[width=0.9\linewidth]{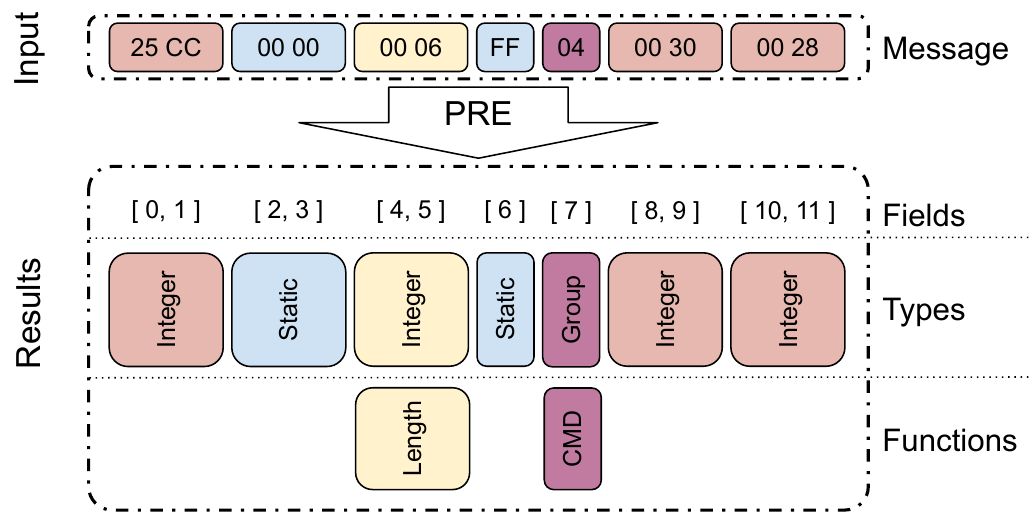}
    \caption{Protocol Reverse Engineering}
    \label{fig:pre}
\end{figure}

Protocol reverse engineering (PRE) is the process of analyzing communication flows to deduce the message structures, semantics, or state machines of network protocols, typically when official specifications are unavailable, incomplete, or proprietary~\cite{narayan2015survey,jiang2024binpre}. When only protocol messages and the program binaries that implement the protocol are available, PRE becomes particularly challenging. Analysts often have to rely on captured network traffic and the behavior of related (often closed-source) software to infer how systems exchange information. Figure~\ref{fig:pre} illustrates the expected output of PRE on a real-world Modbus message~\cite{modbus}. Given a 12-byte message (byte [0] to byte [11]), PRE aims to: (1) accurately identify the boundaries of all fields within the message (field detection), and (2) infer each field’s type and role (semantic inference). PRE is a key enabler for security applications, including analyzing malware command-and-control (C2) communications~\cite{icics23wang,cccs2024fu}, discovering vulnerabilities in undocumented or proprietary protocols~\cite{brumley2007towards}, and generating intrusion detection system (IDS) signatures for previously unknown threats~\cite{dreger2006dynamic}.

Due to the importance of PRE, a substantial body of research has emerged. Existing studies generally fall into two categories: network-traffic-based approaches and binary-analysis-based approaches.
Network-traffic-based methods, such as BinaryInferno~\cite{chandler2023binaryinferno}, NEMESYS~\cite{kleber2018nemesys}, DynPRE~\cite{luo2024dynpre}, and NetPlier~\cite{ye2021netplier}, take captured network traces as input and apply statistical techniques to extract recurring format patterns. However, their effectiveness depends heavily on access to high-quality and diverse traffic that captures many protocol message instances—an assumption that is often difficult to satisfy in practice.
In contrast, binary-analysis-based approaches, including Dispatcher~\cite{caballero2009dispatcher}, PolyGlot~\cite{caballero2007polyglot}, Tupni~\cite{cui2008tupni}, and BinPRE~\cite{jiang2024binpre}, leverage dynamic program analysis to recover detailed runtime semantics by observing how the protocol implementation parses and consumes message bytes.
For example, if the binary consistently loads bytes [0--1] together, applies endian-aware integer operations (e.g., shifting/combining), and uses the resulting value in conditional branches or length calculations, this provides strong evidence that [0--1] forms a multi-byte integer field rather than unrelated bytes.
As a result, binary-based approaches are often more practical in real-world settings where high-quality traffic traces are unavailable.
Despite significant progress, we identify three fundamental limitations of state-of-the-art (SOTA) binary-analysis-based techniques, as elaborated below.

First, SOTA approaches rely heavily on \emph{hand-crafted heuristics} or \emph{rule sets} that are tuned to specific byte patterns or protocol families. This rule-driven design limits generalizability and makes it difficult to keep pace with the diversity of real-world encodings. A representative example is BinPRE~\cite{jiang2024binpre}, which uses multiple rules to label a field as a filename only when the byte sequence ends with a known extension (e.g., ``.txt'' or ``.jpg''). Consequently, filenames that are length-prefixed, UTF-16 encoded, or extensionless can be missed or mislabeled.

Second, existing techniques often \emph{model fields in isolation}, without explicitly capturing semantic relationships among neighboring fields. In practice, many protocols exhibit strong contextual dependencies that are crucial for semantic inference. For instance, a tag immediately followed by a length field strongly suggests the canonical Type--Length--Value (TLV) pattern, and checksum fields typically appear after the data they protect. The Dynamic Host Configuration Protocol (DHCP)~\cite{dhcp}, for example, encodes options as a sequence of TLV triples. Reliably identifying the Tag--Length prefix is essential for delineating the corresponding value fields and inferring their types. Prior work, such as BinaryInferno~\cite{chandler2023binaryinferno}, can leverage serialization patterns to detect certain LV/TLV-like structures, but it does not reason about \emph{why} fields depend on one another or how upstream boundary errors propagate to downstream semantic inference. As a result, failing to model such dependencies can lead to inaccuracies that may cascade into subsequent analyses, such as vulnerability discovery.

Third, these PRE techniques struggle to robustly handle a special yet prevalent case -- \emph{nested protocols}, where a message embeds a sub-payload that follows a different protocol specification. While dynamic binary-analysis--based approaches (e.g., Polyglot~\cite{caballero2007polyglot}, Tupni~\cite{cui2008tupni}, and BinPRE~\cite{jiang2024binpre}) can expose certain field semantics by observing runtime parsing behavior or can infer repeated or nested layouts from control-flow regularities, these methods primarily recover syntactic structure rather than an explicit \emph{hierarchical} semantic representation. In practice, inferred fields are often flattened into a single sequence, obscuring protocol-layer boundaries. However, nested payloads typically induce stronger dependencies \emph{within} a sub-protocol than \emph{across} protocol boundaries. Existing systems do not explicitly represent or reason about this distinction, and approaches with limited semantic cues (e.g., BinaryInferno) still assign roles largely at the level of individual fields, without modeling hierarchical relationships or distinguishing intra-payload dependencies from inter-layer transitions. Consequently, nested protocols may be only partially recovered, and inferred layouts can become brittle for multi-layer designs such as EtherNet/IP (EIP)~\cite{brooks2001ethernet} encapsulating Common Industrial Protocol (CIP)~\cite{viktor2016common} payloads.

Recent studies~\cite{wong2025decllm,hu2024degpt,she2024wadec} suggest that large language models (LLMs) can outperform rigid, heuristic-based pipelines in reverse-engineering tasks such as disassembly~\cite{she2024wadec} and decompilation~\cite{hu2024degpt,wong2025decllm}. Trained on large corpora, LLMs can internalize recurring structural regularities and, importantly, capture relationships among distant components, enabling more flexible inference under uncertainty. Motivated by these advances, we first examined whether an off-the-shelf LLM could perform protocol reverse engineering (PRE) without additional structure. In a pilot study, we asked GPT-o3 (the same backbone model used in our approach) to perform field detection on 179 messages from four representative protocols (50 Modbus, 41 S7, 38 FTP, and 50 DNS). The results exposed the limitations of naïve prompting: the model achieved an F1 of 0.69 on Modbus (the simplest format in our set) and remained below 0.50 on S7, FTP, and DNS. These outcomes indicate that dependable PRE requires more than a single, unconstrained pass through an LLM. A qualitative inspection suggests that naïve prompts often trigger hallucinated fields, incorrect semantic assumptions, and inconsistent labeling across messages, especially for formats with optional fields, variable-length encodings, or layered structure. We attribute this behavior to the difficulty of reliably grounding generation on long, noisy inputs (e.g., dynamic traces) and maintaining global consistency when many interacting constraints must be satisfied simultaneously.

To effectively tackle the PRE problem, our key insight is that \emph{field detection and semantic inference are tightly coupled}: accurate boundaries make semantics easier to infer, while semantic constraints (e.g., TLV relationships, checksums, length fields, nested payload headers) can in turn validate and refine boundaries. With this, we develop \codename, a fully automated PRE framework, which integrates fine-grained dynamic program analysis with a recursive LLM-driven reasoning procedure that performs \emph{joint} optimizations over field detection and semantic inference, rather than treating them as two independent stages as in prior work. \codename first performs fine-grained dynamic taint analysis and loop detection while executing the protocol binary on a given input message. These traces provide low-level evidence about how individual bytes are consumed (e.g., grouped loads, comparisons, arithmetic, and repeated parsing behavior), which we encode as structured hints for downstream reasoning. Next, in the field detection step, \codename uses an LLM-based tree-of-thought search to propose candidate segmentations that are consistent with the collected taint/loop evidence. To address nested protocols, \codename exploits the LLM’s capability to recognize sub-payload boundaries and recursively applies the same detection procedure to any inferred nested regions, yielding an initial \emph{hierarchical} message layout. Given this layout, our system performs semantic inference to predict the type and role of each field. The inferred semantics are then fed back into a subsequent detection round: high-confidence boundaries are frozen, and ambiguous regions are re-segmented under stronger semantic constraints. Then, \codename alternates between field detection and semantic inference, using the output of each step to constrain the other until a fixed point is reached. If the process fails to converge within a preset budget, \codename restarts with alternative hypotheses to escape poor local optima. Inspired by the Expectation--Maximization (EM) algorithm~\cite{EM_algorithm}, we formalize this procedure as an iterative optimization process and prove that each update is monotonic and that the procedure converges to a local optimum. The end result is a heuristic-free, effective PRE solution that improves both accuracy and stability compared to state-of-the-art techniques.

We implemented a prototype of \codename and evaluated it against six state-of-the-art PRE tools. The results demonstrate that \codename achieves the best overall performance across a diverse set of popular protocols.
In field detection, \codename attains the highest average F1 score of 0.76, demonstrating a 90\%+ improvement with no other tools exceeding 0.4.
In semantic inference, \codename achieves an average F1 score of 0.72, substantially outperforming the 0.14 of BinPRE, the state-of-the-art PRE tool.
Our further ablation study confirms and quantifies the contribution of each design component.
In addition, we demonstrate \codename's practicality for security tasks by using it to analyze the protocol of a real-world, widely distributed botnet malware sample.

\vspace{4pt} \noindent \textbf{Contributions.} Our contributions are summarized as follows:
\begin{itemize}
    \item We designed a fully automated PRE framework, which integrates fine-grained dynamic program analysis with a recursive LLM-driven reasoning procedure that performs \emph{joint} optimizations over field detection and semantic inference.
    \item We formally proved that our algorithm converges to a local optimum. 
    \item We implemented a prototype of \codename and open-sourced our implementation~\footnote{https://anonymous.4open.science/r/SemPRE-F9F8} to facilitate future research in this field. We further evaluated its efficacy. Our experiments demonstrate that \codename significantly outperforms prior PRE mechanisms and excels in analyzing real-world malware protocols (\S\ref{sec:eval}).

\end{itemize}

\section{Problem Statement \& Motivation}

\subsection{Problem Statement}

We formally define the PRE problem as follows. Let $\mathcal{M}$ denote the space of protocol messages. Each message
$m \in \mathcal{M}$ is a finite byte sequence
\vspace{-2mm}
$$
m = (b_0, b_1, \dots, b_{|m|-1}), \quad b_i \in \{0,1\}^8.
$$

\vspace{4pt}\noindent\textbf{Field detection.}
A (flat) segmentation of $m$ is an ordered list of $K$ byte intervals

$$
s = \big((l_1,r_1), (l_2,r_2), \dots, (l_K,r_K)\big)
$$
such that (i) each interval is contiguous with $0 \le l_k \le r_k \le |m|-1$,
(ii) intervals are non-overlapping and ordered, i.e., $r_k + 1 = l_{k+1}$ for all $k \in \{1,\dots,K-1\}$, and
(iii) they cover the entire message, i.e., $l_1=0$ and $r_K=|m|-1$.
Let $\mathcal{S}(m)$ denote the set of all valid segmentations of $m$.
Field detection seeks a segmentation that maximizes a global structural objective:
$$
f_{\mathrm{det}}(m) = \arg\max_{s \in \mathcal{S}(m)} J(m, s),
$$
where $J(m,s)$ scores structural plausibility (e.g., boundary quality and cross-field constraints).

\vspace{4pt}\noindent\textbf{Semantic inference.}
Let $\mathcal{Y}_{\mathrm{common}}$ be a vocabulary of semantic labels (field types/functions), and let
$$
\mathcal{Y} = \mathcal{Y}_{\mathrm{common}} \cup \{\bot\},
$$
where $\bot$ denotes ``unknown / none of the above.'' Given a message $m$ and a detected segmentation
$s$ with $K$ fields, semantic inference assigns one label per field by maximizing a global semantic objective:
$$
f_{\mathrm{sem}}(m, s) = \arg\max_{y \in \mathcal{Y}^K} G(m, s, y),
$$
where $y = (y_1,\dots,y_K)$ and $G(m,s,y)$ evaluates semantic consistency across fields and with the observed bytes.
For each field $k$, $y_k \in \mathcal{Y}_{\mathrm{common}}$ indicates an assigned semantic type/function, while $y_k=\bot$ indicates that no suitable label is assigned.

In this study, we reuse $\mathcal{Y}_{\mathrm{common}}$ defined in BinPRE~\cite{jiang2024binpre} (i.e., five common types: static, integer, group, bytes, and string, and six common semantic functions: command, length, delim, checksum, filename, and aligned, detailed in Table~\ref{tab:semantic-types-functions}) for fair comparisons in our evaluation. Note that adding new types or functions is straightforward and requires minimal configuration in \codename, since it is built on an LLM that retains extensive knowledge of types and functions.


\subsection{Motivation}
\label{sec:motivation}
To better understand the limitations of current PRE techniques, we examine several representative scenarios drawn from real-world network protocols and highlight the core challenges that existing tools struggle to address. 

\begin{figure}[h]
    \centering
    \vspace{-2mm}
    \includegraphics[width=\linewidth]{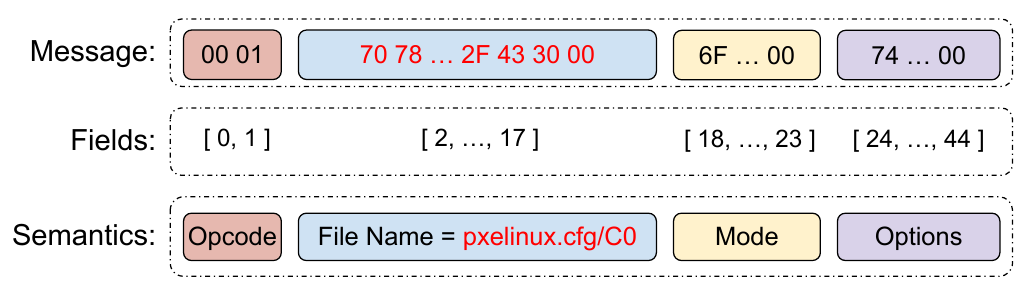}
    \caption{Inflexible Heuristics-Based Rules}
    \label{fig:tftp}
\end{figure}

\vspace{4pt}
\noindent \textbf{Inflexible Heuristics-Based Rules.}
Recent PRE mechanisms (e.g., BinPRE~\cite{jiang2024binpre}) rely heavily on hand-crafted inference rules, typically designed by experts based on well-documented protocols and prior knowledge, to identify and label protocol fields. These rules inevitably introduce bias and limit the potential for generalization. 
For instance, Figure~\ref{fig:tftp} shows a real-world TFTP message containing a filename field (Byte[2-17]).
The file name here (``pxelinux.cfg/C0'') includes a special slash character that is explicitly excluded by BinPRE's filename identification heuristic,
but permitted by TFTP's specification~\cite{tftp-rfc}.
Consequently, BinPRE fails to correctly recognize field semantics because of its rigid rules. 

In contrast, \codename avoids such brittle, protocol-specific heuristics by grounding an LLM’s semantic hypotheses in dynamic execution evidence (e.g., tainting information and parsing behavior), enabling it to generalize to unseen but spec-compliant field values rather than rejecting them due to hard-coded pattern rules.


\begin{figure}[h]
    \centering
    \includegraphics[width=\linewidth]{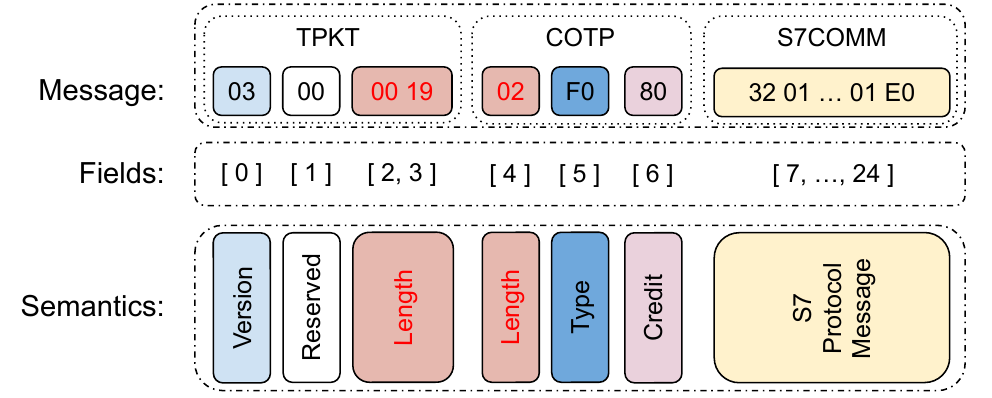}
    \caption{Ignorance of Semantic Relationships}
    \label{fig:s7}
\end{figure}

\vspace{4pt}\noindent \textbf{Ignorance of Semantic Relationships.}
Protocol fields rarely occur in isolation; instead, they are tied together by structural and semantic dependencies that govern correct parsing and interpretation. A common example is the \emph{Length--Value (LV)} construct, in which a length field specifies the size of a subsequent variable-length field, allowing the receiver to precisely delimit the payload. Figure~\ref{fig:s7} shows such dependencies in an S7 communication message~\cite{s7-spec}. In this example, byte[4] encodes a length that indicates the next 2 bytes form a payload segment. Likewise, byte[2--3] encodes the total message length, constraining the maximum extent of all subsequent fields. These length declarations create explicit cross-field constraints that, if modeled, can substantially reduce ambiguity in both field boundary detection and semantic labeling.

Existing PRE techniques do not reason about these inter-field dependencies, which makes them prone to cascading errors. For instance, a boundary mistake around a length field can distort the inferred layout of all downstream fields, and semantic mislabeling can remove key constraints needed for correct segmentation. For instance, BinPRE infers semantics at the byte (or local) level and labels byte[4] as a fixed-width numeric field without linking it to the following payload, and it even misinterprets byte[2--3] as two independent ``Length'' fields rather than a single two-byte length value. \codename addresses this limitation by \emph{jointly} optimizing field boundaries and semantic labels in an iterative loop, allowing inferred LV constraints (e.g., a length field governing a following payload) to validate and refine segmentation decisions rather than treating each field in isolation.

\begin{figure}[h]
    \centering
    \includegraphics[width=\linewidth]{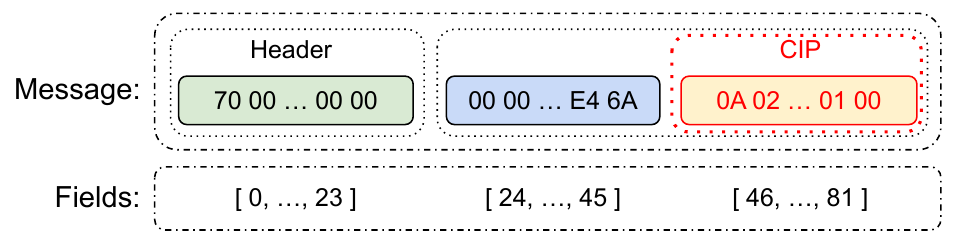}
    \caption{Insufficient Handling of Nested Protocols}
    \label{fig:eip}
\end{figure}

\vspace{4pt}
\noindent \textbf{Insufficient Handling of Nested Protocols.}
Nested protocol structures are pervasive in real-world communication stacks, especially in industrial settings. A representative example is EtherNet/IP (EIP)~\cite{brooks2001ethernet}, which commonly encapsulates Common Industrial Protocol (CIP)~\cite{viktor2016common} payloads. In practice, a single EIP message may contain one or more embedded CIP messages, each with its own internal field layout and semantics. As shown in Figure~\ref{fig:eip}, the example EIP message includes a CIP sub-payload spanning byte[46--81], accounting for nearly half of the message.

Current techniques implicitly assume a single, homogeneous message format. Hence, they either stop at the outer-layer header/payload boundary (leaving the embedded protocol unanalyzed) or flatten the entire byte stream into a single structure, which often leads to missegmentation and incorrect semantic labels for the inner protocol. \codename explicitly targets nested protocols via \emph{recursive} inference. 


\section{System Design and Implementation}

\subsection{Overview}\label{sec:overview}

\begin{figure}[h]
    \centering
    \includegraphics[width=\linewidth]{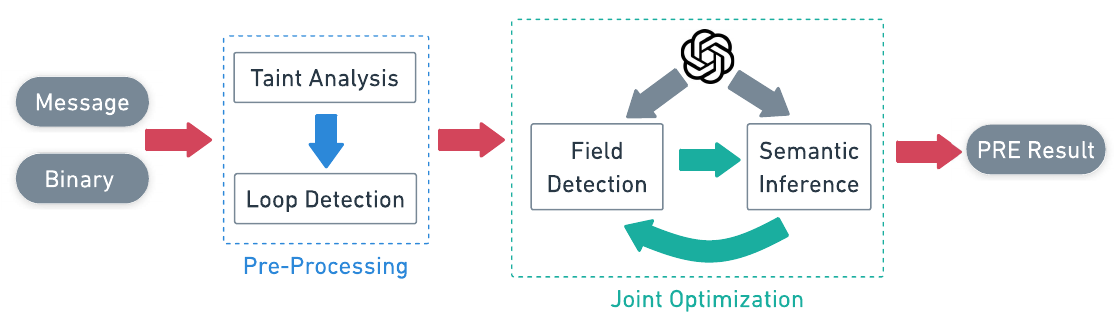}
    \caption{\codename Overview}
    \label{fig:overview}
\end{figure}

As depicted in Figure~\ref{fig:overview}, \codename comprises three principal phases, namely pre-processing, field detection, and semantic inference, of which the latter two are coupled through a joint optimization process.

Given a raw message from an undocumented protocol and its corresponding binary, preprocessing runs the binary on the message under dynamic taint analysis. We record per-byte dataflow and detect parsing loops along the executed path, exposing where the program repeatedly consumes input. These taint/loop cues reliably indicate variable-length arrays and length-prefixed regions, helping localize field boundaries and size dependencies. The taint map, loop inventory, and raw message are then used to construct an initial segmentation by aligning taint boundaries with loop patterns and other structural cues. Next, semantic inference assigns each segment a data type and functional role (e.g., length, command, string) using evidence extracted from the binary and statistical priors. High-confidence semantic predictions are fed back as constraints to refine segmentation (e.g., enforcing length–payload consistency). Field detection and semantic inference alternate until the joint objective stops improving; if needed, we restart with a different initialization up to a fixed budget. The final output is a stable, semantically annotated message layout. Notably, we operate on each message independently and avoid cross-message semantic alignment or majority-vote reconciliation. Without ground truth, global alignment can amplify noisy predictions and propagate errors across messages; instead, \codename emphasizes per-message internal consistency guided by execution evidence.

This process mirrors the workflow of a human analyst: an initial coarse segmentation and semantic assignment is iteratively refined by identifying high-confidence labels and using them to revise earlier boundary and type hypotheses, repeating until the message description becomes coherent and self-consistent. In \codename, the analyst’s evaluative role is delegated to an LLM. Recent studies~\cite{madaan2023self, hu2024teaching, lu2024self, liang2024internal} show that state-of-the-art models (e.g., GPT-4) can perform self-critique and error correction when provided access to intermediate reasoning artifacts, and prior work~\cite{chiang2023can, sun2023principle, fu2023gptscore} finds their judgments often track human preferences, making them effective automated evaluators in iterative pipelines~\cite{achiam2023gpt}. Accordingly, after each iteration, the LLM reviews the current hypotheses, flags outputs it deems reliable, and feeds these constraints into the next cycle to progressively improve both segmentation and semantic interpretation.


\subsection{Pre-processing}
\label{sec:preprocess}

In the pre-processing stage, given a message and a binary that consumes the message, our system performs a byte-level dynamic taint analysis~\cite{schwartz2010all} that taints the given message as the data source and keeps track of how the binary processes the given message during the execution by recording how the binary's instructions access, propagate, and transform the tainted data from the input message. By doing so, \codename can produce a set of taint information that contains details on which instructions read from, write to, or otherwise depend on each part of the input message during execution, which is essential for further segmenting the message into multiple fields. This information provides execution-level evidence for subsequent field segmentation. Since \codename performs semantic reasoning on a per-message basis, its effectiveness does not primarily depend on the number of execution traces collected for the same message. To further facilitate field detection, we perform loop detection~\cite{moseley2006loopprof} to recognize loops during the execution and record the memory addresses of the loops in order to correlate repetitive processing patterns observed in the execution trace with the specific message bytes being handled. Identifying loops will highlight code that iterates over portions of the input, revealing variable-length fields or length-prefixed blocks. Such iteration patterns offer clear markers for field boundaries and size relationships, thereby sharpening the accuracy.

\subsection{Field Detection}
\label{sec:field-detection}

\begin{figure}[h]
    \centering
    \includegraphics[width=\linewidth]{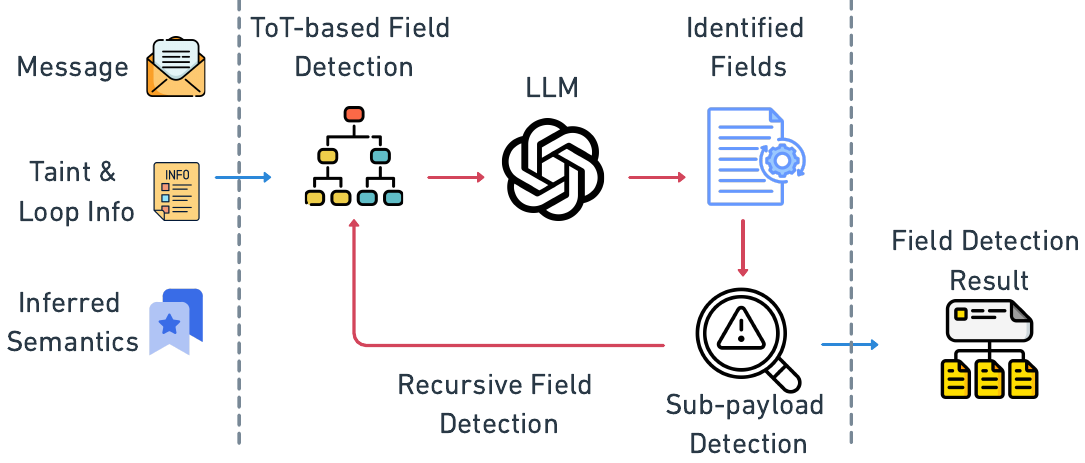}
    \caption{Field Detection}
    \label{fig:field}
\end{figure}

As depicted in Figure~\ref{fig:field}, our field detection takes as input the message, the taint and loop information from the pre-processing, and the inferred semantics from semantic inference (will be empty in the first iteration, more details in Section~\ref{sec:semantics}), and eventually outputs the field detection result. To achieve this, \codename leverages a novel prompting technique, Tree-of-Thoughts (ToT)~\cite{yao2023tree}, which enables LLMs to explore multiple reasoning paths through thoughts, leading to more accurate problem-solving. Specifically, ToT in field detection enables the LLM to cross-validate multiple candidate segmentations against taint, loop, and semantic cues, thereby pruning inconsistent branches and curbing hallucinations while arriving at a globally coherent field layout, which effectively improves accuracy.


\vspace{4pt}\noindent \textbf{Tree-of-Thoughts (ToT) in Field Detection.} ToT is a prompting framework that extends chain-of-thought prompting and self-consistency~\cite{wang2022self,wei2022chain}. Instead of committing to a single linear reasoning trace, ToT explores multiple candidate reasoning trajectories in parallel, organized as branches of a tree. Each node corresponds to a \emph{thought}: a short textual intermediate step that is sufficiently concise to enable diverse generation, yet informative enough to support reliable evaluation. A typical ToT workflow comprises four steps: (1) \emph{thought decomposition}, which defines the intermediate steps appropriate for the task; (2) \emph{thought generation}, which proposes $k$ candidate thoughts for the current state (e.g., via sampling); (3) \emph{state evaluation}, which scores candidates by how much progress they make toward a solution and selects the most promising states to retain; and (4) \emph{search}, which expands the tree using a strategy such as breadth-first or depth-first search, iterating until a satisfactory solution is found.

\begin{figure}[h]
    \centering
    \includegraphics[width=\linewidth]{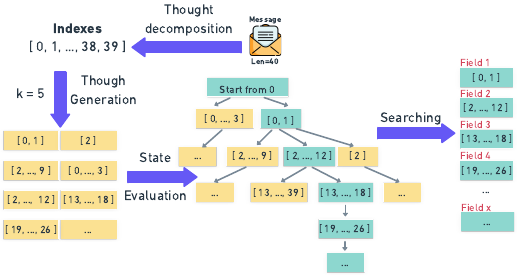}
    \caption{Tree-of-Thoughts Process}
    \label{fig:tot}
\end{figure}


To use ToT effectively in the context of \codename, we instantiate it as a structured search procedure for message segmentation that explicitly captures semantic dependencies between adjacent fields. We treat \emph{field boundaries} as the ToT search space: a ToT \emph{state} represents a partial segmentation of the message prefix, and each expansion hypothesizes the next contiguous field starting at the first unassigned byte. Given a raw protocol message of length $L$, the initial state spans indices $[0,1,\dots,L-1]$ and the search begins at index $i=0$. Because protocol fields are contiguous and sequential, we map one ``thought'' to one complete field. Concretely, at position $i$, a thought proposes the next field as a contiguous byte range $[i,j]$ (with $j \ge i$). Selecting $j$ uniquely determines the next start position $i \leftarrow j+1$, so later boundary decisions are naturally conditioned on earlier ones.

In the thought generation phase, starting from the current index $i$, \codename proposes up to $k$ (we empirically set to 5) end positions $j$, yielding $k$ alternative field hypotheses $[i,j]$. These candidates are not arbitrary: they are generated by jointly leveraging dynamic taint cues (e.g., grouping bytes consumed by the same semantic operation) and LLM-derived protocol cues (e.g., delimiter patterns, length constraints, and value formats). Figure~\ref{fig:tot} illustrates
this process on a message of length $L=40$ (indices $[0,\dots,39]$): at $i=0$, the model may propose candidate boundaries such as $[0,1]$ and $[0,\ldots,3]$; after choosing $[0,1]$, the next step begins at $i=2$ and may generate candidates including $[2,2]$, $[2,\ldots,9]$, and $[2,\ldots,12]$.

In the state evaluation phase, each candidate field induces a successor state (i.e., a longer prefix segmentation), which the LLM scores using a cost-based function, which reflects: (1) whether tainted bytes used in semantic operations are consistently covered by the proposed fields, (2) whether boundaries violate inferred structural/semantic constraints (e.g., length--payload consistency), and (3) whether the partial layout is plausible under common protocol patterns. The cost value $C$ lies in [0, 1] for every path, with $C = 0$ denoting a perfectly consistent segmentation and $C = 1$ denoting one that violates every constraint on every field. \codename retains only low-cost states, pruning implausible segmentations early while preserving diverse, high-quality boundary hypotheses.

Finally, in the search phase, \codename performs a tree search over these states. Starting at $i=0$, we apply depth-first search to enumerate complete segmentations and select the lowest-cost root-to-leaf path that reaches the end of the message. In Figure~\ref{fig:tot}, the final field-detection result is the concatenation of the fields along this best path (i.e., Field~1: $[0,1]$, Field~2: $[2,\ldots,12]$, Field~3: $[13,\ldots,18]$, Field~4: $[19,\ldots,26]$).

\vspace{4pt}\noindent \textbf{Recursive Sub-payload Detection.} We leverage the LLM to perform sub-payload detection, shown in Figure \ref{fig:field}. When the LLM identifies a potential sub-payload region inside a protocol data unit (PDU), we \emph{recursively} invoke the ToT-based field detection procedure described earlier on that sub-region, providing the same inputs (e.g., taint-flow and loop evidence, as well as any currently inferred semantics). This recursive process continues until the LLM reports that no further sub-payloads remain. 
\begin{algorithm}[]
\small
\onehalfspacing
\caption{Field Detection}\label{alg:field_detection}
    \SetKwInOut{KwIn}{Input}
    \SetKwInOut{KwOut}{Output}
    \SetKwFunction{kwFn}{FieldDetection}
    \kwFn{$M, Taint, Loop, SemLabels$}\;
    \KwIn{$M$ - A message of an unknown protocol;\\ $Taint$ - Taint information;\\ $Loop$ - Loop information;\\  $SemLabels$ - High-confident semantic labels; \\ $ToT$ - ToT parameters.}
    \KwOut{$Fields$ - Detected fields of the message;}

    $Prompt_{field} \gets$ Create a field detection prompt using ($Msg, Taint, Loop, SemLabels, ToT$) \;
    $Response \gets$ LLMQuery($Prompt_{field}$) \;
    $Fields \gets Responce$\;
    $Log \gets Prompt_{field}$\;
    \While{i $\leq$ MaxIterTimes}{
        $Prompt_{field} \gets Log + Fields + P_{subpayload}$\;
        $Response \gets$ LLMQuery($Prompt_{field}$)\;
        $Fields, SubPayloadFlag \gets Response$\;
        \eIf{SubPayloadFlag $==$ True}{
         $Log \gets Prompt_{field}$\;
         $i++$\;
         continue\;
         }{
         break\;
        }
    }
    return $Fields$
\end{algorithm}

As illustrated in Algorithm~\ref{alg:field_detection}, it takes as input a message $M$, the taint and loop information $Taint$ and $Loop$ from pre-processing, and high-confident semantic labels $SemLabels$ from semantic inference (which will be empty in the first iteration), and outputs $Fields$ as detection results. Specifically, we first create a field-detection prompt $Prompt_{field}$ using the given parameters (Ln. 2) and query the LLM to obtain $Response$, which contains the initial field set $Fields$. Next, we build a sub-payload detection prompt by combining contextual information $Log + Fields$ with sub-payload prompt template $P_{subpayload}$ (Ln.7) and query LLM to get $Response$ (Ln. 8), which provides updated fields of the message and an indicator showing whether additional sub-payload detection is required. Taint information in the context informs LLM which bytes are fixed, allowing it to reason only over untainted regions. If another iteration is needed, then we update $Log$ and repeat; otherwise, the fields are the results.




\vspace{-3mm}

\subsection{Semantic Inference}~\label{sec:semantics}
\vspace{-2mm}

\begin{figure}[h]
    \centering
    \includegraphics[width=\linewidth]{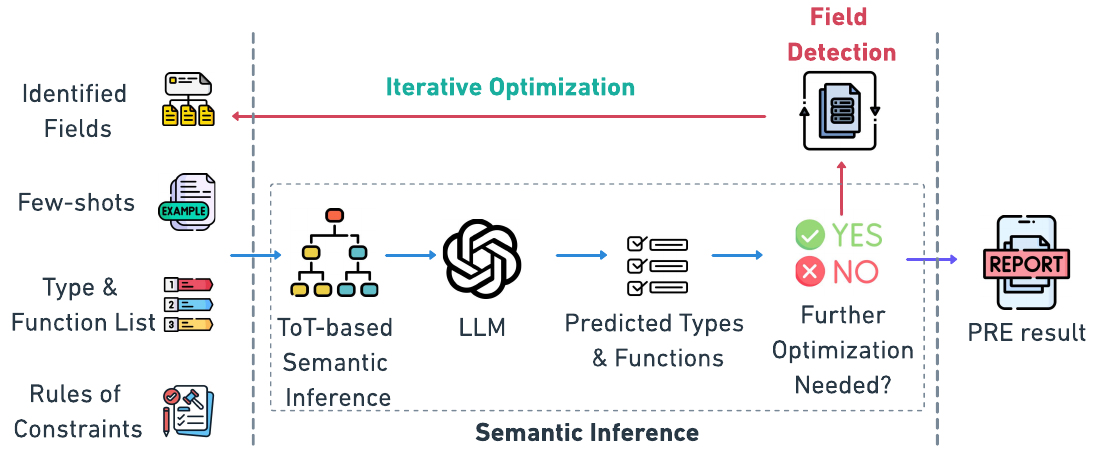}
    \caption{Semantic Inference and Joint Optimization}
    \label{fig:semantic}
\end{figure}

Following field detection, \codename proceeds to semantic inference depicted in Figure~\ref{fig:semantic}. The inference stage is initialized with two resources: (i) a diverse exemplar set covering various protocol categories (e.g., binary, textual, or mixed messages), and (ii) the catalog of candidate type–function pairs defined by BinPRE \cite{jiang2024binpre}. Because a field’s interpretation depends on neighboring fields and message-level constraints, semantic inference cannot be performed as an independent label assignment. For instance, a single field often admits several plausible type–function hypotheses whose validity is contingent on adjacent semantics (e.g., a putative length indicator dictates how the succeeding payload must be parsed).

To this end, we formulate semantic inference as a structured search and realize it with another Tree-of-Thoughts (ToT) prompting scheme. Each field produced by the field-detection phase constitutes one “thought” step: at that node, the LLM generates candidate type–function pairs by analyzing byte patterns, taint-/loop-derived behavior, and higher-level protocol cues. Successive steps branch to form alternative semantic paths over the message. Every path is cost-scored according to constraint consistency and corpus priors; a branch-and-bound mechanism then prunes incoherent paths, retaining only those with minimal cost. Exploration proceeds in sliding windows of 5 fields, iterating until all segments have been evaluated. During the process, the LLM also serves as an internal evaluator. Paths whose cost falls below a calibrated threshold are marked as high-confidence, and their labels are fixed as additional constraints for the next iteration. Through this iterative coupling of field detection and semantic inference, \codename progressively converges on a globally coherent structural and semantic interpretation of the message. Complete prompt templates are included in Appendix~\ref{sec:prompt_tmp}.





\subsection{Joint Optimization}
\label{sec:iter-refine}
Based on our unique insight that field detection and semantic inference are strongly correlated and can complement each other, \codename performs an iterative process between the two to jointly optimize results and reach a local optimum, as shown in Figure~\ref{fig:semantic}. This component is inspired by the Expectation–Maximization (EM) algorithm~\cite{EM_algorithm}, which addresses problems involving partially observed data. In particular, after the initial inference round, \codename has identified message fields and predicted their types and functions. Among these, some predictions can be marked as high-confidence (evaluated by the LLM) and serve as partially observed labels that guide the next iteration.
This process is also inspired by how human analysts approach reverse engineering tasks. Analysts often begin with uncertain or approximate field interpretations, refine their understanding as more structural or semantic cues emerge, and use confident insights to correct earlier errors. \codename mimics this behavior by using the LLM to evaluate its own intermediate outputs and determine whether further refinement is warranted.
Through this feedback loop, \codename incrementally improves its predictions, correcting earlier mistakes and reinforcing accurate interpretations. The refinement process continues until semantic convergence is achieved or no further improvements can be made.

\begin{algorithm}[!htbp]
\small
\onehalfspacing
\caption{Joint Optimization}\label{alg:optimization}

    $Fields = \{\}, SemLabels = \{\}$\; 
    \While{i $\leq$ MaxIterationTimes}{
        $Fields \gets$ FieldDetection$(M, Taint, Loop, SemLabels) $\; 
        $SemLabels \gets$ SemanticInference$(M, FewShots, Fields) $\; 
        $Fields_h, SemLabels_h, NeedFurtherOp \gets $ LLMQuery $(Fields, SemLabels)$\;
        update $(Type_h, Fn_h)$\;
            \eIf{$NeedFurtherOp == True$}{
             $Fields = Fields_h$\;
             $SemLabels$ = $SemLabels_h$\;
             $i++$\;
             Go to the next iteration \;
             }{
             return $Fields, SemLabels$\;
            }

    }
    \If{ $i > MaxIterationTimes$}{
        Restart the whole process \;
    }
\end{algorithm}

\noindent\textbf{Optimization Algorithm.} The optimization procedure is summarized in Algorithm~\ref{alg:optimization}. We initialize the field detection results, denoted by $Fields$, and the semantic inference results, denoted by $SemLabels$, as empty sets. The algorithm then proceeds iteratively through the following stages: (1) invoke Algorithm~\ref{alg:field_detection} to obtain a provisional set of fields (Ln.3); (2) apply few-shot exemplars, constraint rules, and the interim fields to infer semantic labels (Ln.4); and (3) query the LLM to select high-confidence predictions and determine whether additional joint optimization is needed (Ln.5). If further optimization is required, the high-confidence predictions are fixed, and the next iteration is executed (Ln.7–11). Otherwise, the algorithm returns $Fields$ and $SemLabels$ as the final outputs (Ln.13). Moreover, if the current solution has not converged, \codename restarts the entire process and repeats the optimization until a predefined limit is reached (Ln.16–17).

\subsection{Analysis of Iterative Refinement}

\codename alternates between field detection and semantic inference, using high-confidence predictions from one iteration as constraints for the next. Because both stages involve stochastic LLM inference, we do not claim unconditional convergence for arbitrary executions. Instead, we analyze the refinement process under a structural property of the PRE problem and two practical behavioral assumptions. The analysis characterizes when the iterative process is expected to stabilize.

\vspace{4pt} \noindent \textbf{State and Objective.}
Given a message $B$ and its dynamic analysis information $T$, let $S_t$ and $L_t$ denote the field segmentation and semantic labels at iteration $t$, respectively. Let $H_t$ denote the set of high-confidence predictions retained as constraints. We use $J(S, L\mid B, T)$ to denote the consistency of a candidate interpretation with runtime evidence and cross-field constraints, where a larger value indicates a more consistent interpretation.

\vspace{4pt}\noindent \textbf{Finite Candidate Space.}
For a finite message, the number of possible contiguous segmentations is finite. Since \codename assigns each field a label from a finite semantic vocabulary, the joint candidate space of $(S, L)$ is also finite. Likewise, $H_t$ can contain only finitely many constraints. This property follows directly from the discrete formulation of field segmentation and semantic labeling.

\vspace{4pt}\noindent \textbf{Assumption 1: Reliable High-Confidence Constraints.}
Following prior findings on the reliability of high-confidence LLM predictions, we assume that predictions retained in $H_t$ are sufficiently reliable to serve as constraints in subsequent iterations. In particular, we assume that fixing these predictions does not introduce erroneous constraints that prevent refinement toward an interpretation more consistent with the available runtime evidence. The retained constraints therefore accumulate as
\begin{equation}
    H_t \subseteq H_{t+1}.
\end{equation}

\vspace{4pt}\noindent \textbf{Assumption 2: Consistency-Preserving Refinement.}
We further assume that an accepted refinement does not decrease the consistency objective:
\begin{equation}
    J(S_{t+1},L_{t+1}\mid B,T)
    \geq
    J(S_t,L_t\mid B,T).
    \label{eq:monotonic}
\end{equation}
This condition captures \codename's intended behavior: newly identified semantic constraints, such as a reliable length--payload relationship, reduce ambiguity in subsequent field detection, while refined field boundaries constrain semantic inference.

Under Eq.~\ref{eq:monotonic}, the sequence
\[
J_t = J(S_t,L_t\mid B,T)
\]
is non-decreasing. Because the candidate space is finite, $J_t$ cannot strictly increase indefinitely. Moreover, if an iteration does not improve $J_t$ but adds new reliable constraints, $H_t$ can strictly grow only finitely many times. Therefore, if \codename continues refinement only when either
\begin{equation}
    J_{t+1} > J_t
    \quad \text{or} \quad
    H_t \subset H_{t+1},
\end{equation}
The process eventually reaches a state in which neither condition holds. We call this a {stable fixed point} of the refinement procedure.

\vspace{4pt}\noindent \textbf{Practical Interpretation.}
The above result is conditional, not a formal guarantee of local or global optimality. In practice, an LLM may generate an inconsistent hypothesis or incorrectly mark a prediction as high-confidence, violating the behavioral assumptions above. We can observe behavior consistent with these assumptions across iterations by tracking the accumulation of high-confidence predictions and changes in the predicted field boundaries and semantic labels. \codename also bounds the number of refinement iterations and restarts the process when it does not stabilize within the predefined budget. 

Thus, the analysis characterizes when \codename's refinement process reaches a stable fixed point, rather than claiming convergence to a local or global optimum.

\section{Evaluation}
\label{sec:eval}
We evaluate our system with respect to its general effectiveness and its efficacy in handling real-world cases. Specifically, we aim to conduct a series of experiments to answer the following research questions. 
\begin{itemize}[leftmargin=*]
    \item \textbf{RQ1}: How effective is \codename for field detection?
    \item \textbf{RQ2}: How effective is \codename for semantic inference?
    \item \textbf{RQ3}: Can \codename maintain its effectiveness without potential prior knowledge from the LLM or with a different backbone LLM?
    \item \textbf{RQ4}: Does each component of \codename contribute uniquely in protocol reverse engineering?
    \item \textbf{RQ5}: Can \codename effectively handle real-world malware protocols?
\end{itemize}

\subsection{Experimental Setup}\label{sec:exp_setup}

\vspace{4pt}\noindent \textbf{Baseline Techniques.} To evaluate the effectiveness of \codename, we compare its performance against six state-of-the-art PRE techniques, including both network-traffic-based and binary-analysis-based approaches. Specifically, we compare \codename with PolyGlot~\cite{caballero2007polyglot}, Tupni~\cite{cui2008tupni}, AutoFormat~\cite{lin2008automatic}, BinPRE~\cite{jiang2024binpre}, BinaryInferno~\cite{chandler2023binaryinferno}, and DynPRE~\cite{luo2024dynpre}. Note that we directly reuse the code released by the original authors for BinPRE, PolyGlot, AutoFormat, and Tupni, and re-implement BinaryInferno and DynPRE ourselves based on their papers since their code is unavailable.

\vspace{4pt}\noindent \textbf{Dataset.}
We constructed the dataset as follows. 

\begin{itemize}[leftmargin=*]
    \item \noindent\textit{Protocols.} We collect three protocol datasets for evaluation: \textbf{D1}) the dataset in BinPRE~\cite{jiang2024binpre}, which includes eight widely used protocols covering textual, binary, and mixed formats across the industrial control and network communication domains; \textbf{D2}) a real-world undocumented dataset from GitHub, which contains three real-world undocumented protocols (KTP~\cite{ktp-source-code}, CCN~\cite{ccn-source-code} and Xfer~\cite{xfer-source-code}); \textbf{D3}) we use Mirai~\cite{mirai-source-code}, which is a botnet malware protocol, as a case study. Further information is listed in Table~\ref{tab:messages}. 
    

    \item \noindent\textit{Messages.} Detailed in Table~\ref{tab:ground_truth_details} in Appendix, we collect two message datasets: (1) open-source message datasets and (2) datasets built by ourselves when open-source datasets are unavailable. Concretely, we build the datasets for DNP3.0, HTTP, Mirai, KTP, CCN, and Xfer using their clients and the \texttt{tcpdump} tool. The other messages are from open-source datasets. The number of messages varies across protocols, as the sources of messages differ by protocol. Additionally, the network-based PRE methods, BinaryInferno and DynPRE, are designed to handle small input messages. Please note: although BinPRE reports results, its public repository does not release the corresponding messages (except Modbus), so we evaluated them using our own captured ones. 
\end{itemize}

\begin{table}[!htb]\centering
\caption{Groud Truth Details.}\label{tab:ground_truth_details}
\addtolength{\tabcolsep}{-2pt}
\resizebox{\linewidth}{!}{
\begin{tabular}{ccrrrrrr}\toprule
Dataset & Protocol &\# Messages &\# Fields &\# Types &\# Fn.s &R.T. &R.F. \\\midrule
\multirow{8}{*}{D1} & Modbus &50 &378 &378 &100 &1 &0.26 \\
&S7comm &41 &798 &526 &372 &0.66 &0.47 \\
&Ethernet/IP &50 &2802 &2346 &1046 &0.84 &0.37 \\
&DNP3 &53 &692 &452 &212 &0.65 &0.31 \\
&DNS &50 &450 &450 &206 &1 &0.46 \\
&FTP &38 &168 &168 &137 &1 &0.82 \\
&TFTP &16 &112 &96 &32 &0.86 &0.29 \\
&HTTP &62 &1414 &1166 &248 &0.82 &0.18 \\\midrule
\multirow{3}{*}{D2}&KTP &9 &54 &54 &9 &1 &0.17 \\
&CCN &8 &64 &64 &24 &1 &0.38 \\
&Xfer &9 &90 &84 &24 &0.93 &0.27 \\\midrule
\multirow{1}{*}{D3} & Mirai &10 &127 &127 &45 &1 &0.35 \\
\midrule
\multirow{3}{*}{D4} & Ethernet/IP &5 & & & & & \\
& BAC &5 & & & &1 & \\
& OPC &5 &80 &80 &26 &1 &0.33 \\
\midrule
Summary & &396 &7149 &5911 &2455 &0.82 &0.35 \\
\bottomrule
\end{tabular}}
\flushleft
R.T. - ratio of labeled types; R.F. - ratio of labeled functions
\end{table}


\noindent \textbf{Ground Truth Collection.}
The primary source of ground-truth annotations is the official protocol documentation. For Mirai, whose protocol lacks a formal specification, we instead derive the annotations through a manual audit of the publicly available source code. The resulting corpus records, for every message, the byte-level field boundaries together with a single type and a single function label for each field.  All annotations are cross-validated against Wireshark packet dissections, yielding a data set we regard as error-free. Type and function labels are absent for a small subset of fields when the available taxonomy proves insufficient or when discrepancies arise between the written specification and the Wireshark interpretation. 


\vspace{4pt}\noindent \textbf{Evaluation Metrics.}
Field detection produces a set of byte offsets that mark the field boundaries. Each predicted offset is classified as either a true positive or a false positive with respect to the reference annotations, enabling the computation of precision (the proportion of predicted boundaries that are correct), recall (the proportion of actual boundaries that are recovered), and their harmonic mean, the F1 score.  Semantic inference, likewise, is evaluated through these three measures: although the task is nominally multi-class, each type or function assignment can be viewed as a binary decision (correct vs. incorrect) relative to the ground truth. Accordingly, we aggregate the per-field outcomes to obtain precision, recall, and F1 for type labels and function labels. Throughout the study, these three metrics (precision, recall, and F1) serve as the basis for our performance assessment. 

Please note that BinPRE uses offset-level metrics for field detection, considering all bytes in a given message, but this is a flaw because it significantly reduces the impact of incorrect boundaries. For example, BinPRE outputs [-1,2,3,5,6] (5 offsets but 4 fields) against a ground truth [-1,1,5,6]. Based on BinPRE's calculation, 3 of 5 offsets match (-1, 5, 6), giving precision 0.6 even though only 1 field is correctly identified (5, 6). Thus, our metrics directly work at the field level. We check whether predicted field boundaries are correct, which is simpler and more direct than BinPRE.

\vspace{-3mm}
\subsection{RQ1: Effectiveness on Field Detection}

\begin{figure}[h]
    \centering
    \vspace{-3mm}
    \includegraphics[width=0.85\linewidth]{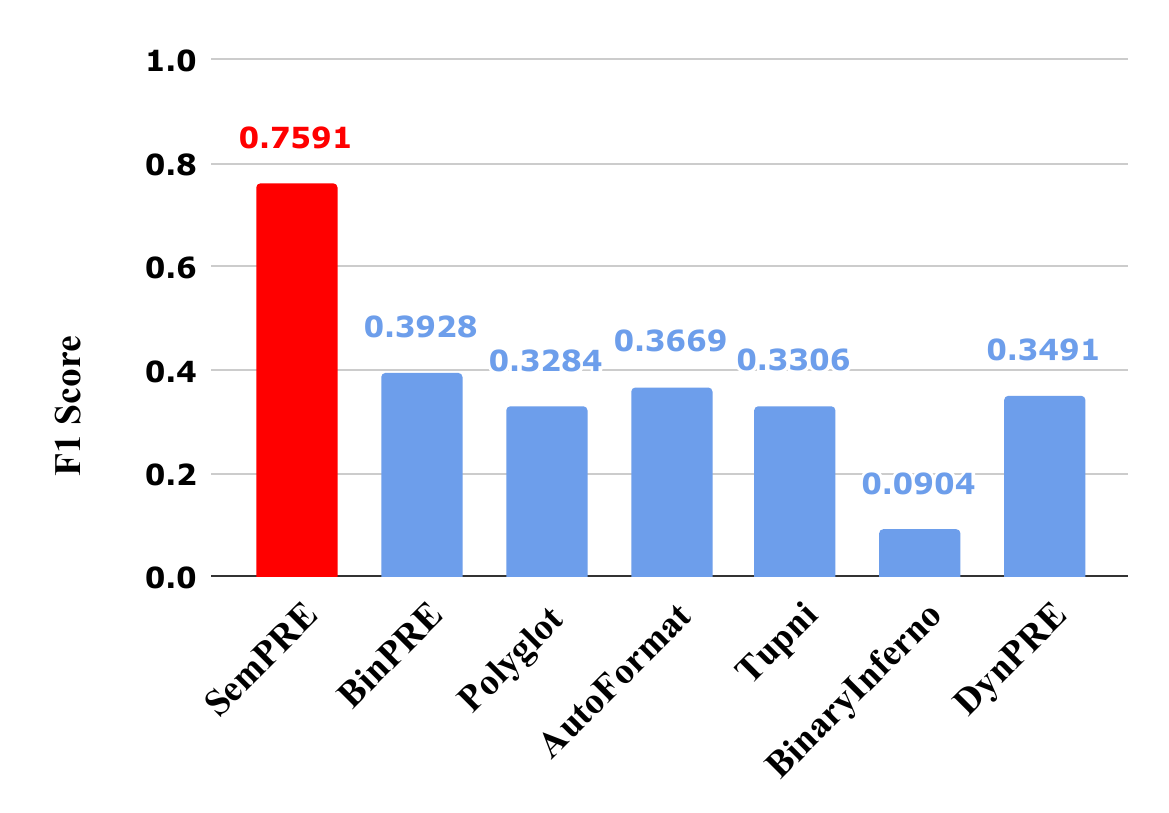}
    \caption{Average F1 Score in Field Detection}
    \vspace{-3mm}
    \label{fig:f1_field_detaction}
\end{figure}

To evaluate field detection, we compute precision, recall, and F1 score for \codename on each protocol and aggregate the results across all messages. We compare against baseline methods using the D1 dataset, which includes eight widely used protocols spanning all three protocol categories. Overall, as shown in Figure~\ref{fig:f1_field_detaction}, \codename substantially outperforms all baselines. Compared to the strongest competitor, BinPRE, \codename achieves a whopping $95\%$ improvement ($0.76$ v.s. $0.39$). Averaged across all baselines, the F1 gain is $0.45$, and remains $0.41$ when excluding the weakest baseline, BinaryInferno. In absolute terms, \codename correctly identifies $4{,}822$ of $6{,}465$ ground-truth fields, exceeding the runner-up by $933$ correct detections.

\begin{table*}[h]
\centering
\caption{Field Detection Results.}\label{tab:field_detection_details}
\addtolength{\tabcolsep}{-2pt}
\scriptsize

\iftrue

\begin{tabular}{rlll|lll|lll|lll|lll|lll|lllr}\toprule
\multirow{2}{*}{Protocols}&\multicolumn{3}{c}{\codename} &\multicolumn{3}{c}{BinPRE} &\multicolumn{3}{c}{Polyglot} &\multicolumn{3}{c}{AutoFormat} &\multicolumn{3}{c}{Tupni} &\multicolumn{3}{c}{Binaryinferno} &\multicolumn{3}{c}{DynPRE} \\\cmidrule{2-22}
 &Pre. &Rec. &F1 &Pre. &Rec. &F1 &Pre. &Rec. &F1 &Pre. &Rec. &F1 &Pre. &Rec. &F1 &Pre. &Rec. &F1 &Pre. &Rec. &F1 \\\midrule
Modbus &0.98 &1.00 &0.99 &0.99 &1.00 &\textbf{1.00} &0.60 &0.98 &0.75 &0.85 &1.00 &0.92 &0.60 &0.98 &0.75 &0.25 &0.15 &0.19 &0.81 &0.37 &0.51 \\
S7comm &0.81 &0.81 &\textbf{0.81} &0.82 &0.02 &0.05 &0.47 &0.16 &0.23 &0.43 &0.13 &0.19 &0.47 &0.16 &0.23 &0.00 &0.00 &0.00 &$\text{0.84}^1$ &$\text{0.51}^1$ &$\text{0.64}^1$ \\
Ethernet/IP &0.86 &0.75 &\textbf{0.80} &1.00 &0.15 &0.25 &0.36 &0.61 &0.45 &0.37 &0.60 &0.46 &0.36 &0.61 &0.45 &0.03 &0.01 &0.01 &0.86 &0.11 &0.19 \\
DNP3 &0.84 &0.76 &0.80 &0.88 &0.58 &0.70 &0.59 &1.00 &0.74 &0.83 &1.00 &\textbf{0.91} &0.69 &1.00 &0.81 &0.98 &0.41 &0.58 &N/A &N/A &N/A \\
DNS &0.56 &1.00 &\textbf{0.72} &0.39 &0.97 &0.56 &0.23 &1.00 &0.38 &0.22 &0.91 &0.36 &0.23 &1.00 &0.38 &0.00 &0.00 &0.00 &$\text{0.21}^2$ &$\text{0.80}^2$ &$\text{0.34}^2$ \\
FTP &0.57 &0.79 &\textbf{0.66} &0.36 &0.38 &0.37 &0.20 &0.58 &0.30 &0.17 &0.44 &0.25 &0.20 &0.58 &0.30 &0.00 &0.00 &0.00 &N/A &N/A &N/A \\
TFTP &1.00 &1.00 &\textbf{1.00} &0.46 &0.66 &0.54 &0.14 &1.00 &0.24 &0.11 &0.83 &0.20 &0.14 &1.00 &0.24 &0.50 &0.17 &0.25 &N/A &N/A &N/A \\
HTTP &0.63 &0.57 &\textbf{0.60} &0.27 &0.36 &0.31 &0.07 &0.40 &0.12 &0.10 &0.40 &0.16 &0.07 &0.40 &0.12 &0.02 &0.00 &0.00 &N/A &N/A &N/A \\
\midrule
Average &0.77 &0.75 &\textbf{0.76} &0.49 &0.33 &0.39 &0.23 &0.60 &0.33 &0.27 &0.58 &0.37 &0.23 &0.60 &0.33 &0.28 &0.05 &0.09 &0.63 &0.24 &0.35 \\
\bottomrule
\end{tabular}
\else
\begin{tabular}{l|rrrrrrrr|r}\toprule
 & Modbus & S7comm & Ethernet/IP & DNP3 & DNS & FTP & TFTP & HTTP & Sum \\\midrule
\multicolumn{10}{c}{\textbf{SemPRE}} \\\midrule
Precision & 0.98 & 0.81 & 0.86 & 0.84 & 0.56 & 0.57 & 0.79 & 0.63 & 0.77 \\
Recall    & 1.00 & 0.81 & 0.75 & 0.76 & 1.00 & 0.79 & 0.79 & 0.57 & 0.75 \\
F1        & 0.99 & \textbf{0.81} & \textbf{0.80} & 0.80 & \textbf{0.72} & \textbf{0.66} & \textbf{0.79} & \textbf{0.60} & \textbf{0.76} \\
\midrule
\multicolumn{10}{c}{\textbf{BinPRE}} \\\midrule
Precision & 0.99 & 0.82 & 1.00 & 0.88 & 0.39 & 0.36 & 0.46 & 0.27 & 0.49 \\
Recall    & 1.00 & 0.02 & 0.15 & 0.58 & 0.97 & 0.38 & 0.66 & 0.36 & 0.33 \\
F1        & \textbf{1.00} & 0.05 & 0.25 & 0.70 & 0.56 & 0.37 & 0.54 & 0.31 & 0.39 \\
\midrule
\multicolumn{10}{c}{\textbf{Polyglot}} \\\midrule
Precision & 0.60 & 0.47 & 0.36 & 0.59 & 0.23 & 0.20 & 0.14 & 0.07 & 0.23 \\
Recall    & 0.98 & 0.16 & 0.61 & 1.00 & 1.00 & 0.58 & 1.00 & 0.40 & 0.60 \\
F1        & 0.75 & 0.23 & 0.45 & 0.74 & 0.38 & 0.30 & 0.24 & 0.12 & 0.33 \\
\midrule
\multicolumn{10}{c}{\textbf{AutoFormat}} \\\midrule
Precision & 0.85 & 0.43 & 0.37 & 0.83 & 0.22 & 0.17 & 0.11 & 0.10 & 0.27 \\
Recall    & 1.00 & 0.13 & 0.60 & 1.00 & 0.91 & 0.44 & 0.83 & 0.40 & 0.58 \\
F1        & 0.92 & 0.19 & 0.46 & \textbf{0.91} & 0.36 & 0.25 & 0.20 & 0.16 & 0.37 \\
\midrule
\multicolumn{10}{c}{\textbf{Tupni}} \\\midrule
Precision & 0.60 & 0.47 & 0.36 & 0.69 & 0.23 & 0.20 & 0.14 & 0.07 & 0.23 \\
Recall    & 0.98 & 0.16 & 0.61 & 1.00 & 1.00 & 0.58 & 1.00 & 0.40 & 0.60 \\
F1        & 0.75 & 0.23 & 0.45 & 0.81 & 0.38 & 0.30 & 0.24 & 0.12 & 0.33 \\
\midrule
\multicolumn{10}{c}{\textbf{Binaryinferno}} \\\midrule
Precision & 0.25 & 0.00 & 0.03 & 0.98 & 0.00 & 0.00 & 0.50 & 0.02 & 0.28 \\
Recall    & 0.15 & 0.00 & 0.01 & 0.41 & 0.00 & 0.00 & 0.17 & 0.00 & 0.05 \\
F1        & 0.19 & 0.00 & 0.01 & 0.58 & 0.00 & 0.00 & 0.25 & 0.00 & 0.09 \\
\midrule
\multicolumn{10}{c}{\textbf{DynPRE}} \\\midrule
Precision & 0.81 & 0.84 & 0.86 & 0.00 & 0.00 & 0.00 & 0.00 & 0.00 & 0.84 \\
Recall    & 0.37 & 0.51 & 0.11 & 0.00 & 0.00 & 0.00 & 0.00 & 0.00 & 0.13 \\
F1        & 0.51 & 0.64 & 0.19 & 0.00 & 0.00 & 0.00 & 0.00 & 0.00 & 0.22 \\
\bottomrule
\end{tabular}
\fi
\flushleft
1 - / 2 - DynPRE only accepts part of the messages\\
1 - Counting based on the first 31 messages;\quad 2 - Counting based on the first 16 messages.
\end{table*}

For individual protocols, \codename outperforms all baselines in nearly all cases (Table~\ref{tab:field_detection_details}), with two minor exceptions. For Modbus, BinPRE achieves a perfect F1 score ($1.00$), while \codename attains $0.99$, a negligible difference. For DNP3, AutoFormat achieves higher precision due to protocol-specific constraints. DNP3 produces the longest taint- and loop-analysis traces because its implementation validates nearly every byte, which limits \codename to using less than $10\%$ of the analysis data under OpenAI API input-length constraints. AutoFormat, by contrast, can leverage the full analysis. Even so, \codename attains an F1 score of approximately $0.80$ (about $90\%$ of AutoFormat’s performance), demonstrating strong robustness despite the reduced data.


\begin{answerbox}
    \textit{Answer to RQ1:} \codename identifies fields with a solid F1 score of $0.76$ on all protocol messages and can beat all state-of-the-art techniques.
\end{answerbox}


\subsection{RQ2: Effectiveness on Semantic Inference}
To evaluate semantic inference, we assess type inference and function inference separately, comparing only against baselines that support the corresponding task. For type inference, we compare exclusively with BinPRE, the only baseline that predicts field types, using the same type categories for fairness. For function inference, we include all baselines except AutoFormat, which does not perform semantic labeling. Overall, \codename correctly infers $3{,}441$ of $5{,}582$ reference types and $1{,}431$ of $2{,}353$ reference functions.

For type inference (Table~\ref{tab:type_infer_results}), \codename consistently outperforms BinPRE across all evaluated protocols. Its F1 score on TFTP is comparatively lower ($0.3542$), but it still exceeds all baselines. This gap is primarily caused by \emph{semantic ambiguity}: several TFTP fields are encoded as byte strings, yet they semantically represent numeric parameters or composite values, making type assignment difficult even for experienced analysts. BinPRE’s weak results on S7comm, FTP, and HTTP largely stem from inaccurate field detection: when boundaries are wrong, semantic inference is inherently inaccurate. While BinPRE reaches an F1 of $0.70$ on DNP3, it often assigns multiple type labels to the same field. We count such multi-label outputs as incorrect (even if one label matches the ground truth), because this behavior amounts to guessing and can artificially inflate precision (by up to $20\%$) without providing genuine semantic understanding.

\vspace{-3mm}
\begin{table}[H]
\centering
\caption{Type Inference Results.}\label{tab:type_infer_results}
\addtolength{\tabcolsep}{-3pt}
\resizebox{0.85\linewidth}{!}{
\begin{tabular}{rrrr|rrrr}\toprule
\multirow{2}{*}{Protocols}  &\multicolumn{3}{c}{\codename} &\multicolumn{3}{c}{BinPRE} \\\cmidrule{2-7}
 &Precision &Recall &F1 &Precision &Recall &F1 \\\midrule
Modbus &0.82 &0.79 &\textbf{0.81} &0.41 &0.40 &0.41 \\
S7comm &0.83 &0.78 &\textbf{0.80} &0.00 &0.00 &0.00 \\
Ethernet/IP &0.90 &0.60 &\textbf{0.72} &0.33 &0.04 &0.08 \\
DNP3 &0.92 &0.75 &\textbf{0.82} &0.00 &0.00 &0.00 \\
DNS &0.75 &0.67 &\textbf{0.71} &0.58 &0.42 &0.49 \\
FTP &0.95 &0.58 &\textbf{0.72} &0.00 &0.00 &0.00 \\
TFTP &0.35 &0.35 &\textbf{0.35} &1.00 &0.16 &0.27 \\
HTTP &0.90 &0.48 &\textbf{0.62} &0.00 &0.00 &0.00 \\
\midrule
Average &0.86 &0.62 &\textbf{0.72} &0.45 &0.08 &0.14 \\
\bottomrule
\end{tabular}}
\flushleft
\end{table}
\vspace{-3mm}

For function inference (Table~\ref{tab:fn_infer_results} in Appendix), \codename again outperforms all baselines on every protocol, yielding a $90\%$ improvement over the runner-up (BinPRE). Performance on S7comm, EIP, and HTTP is comparatively lower ($0.64$, $0.67$, and $0.70$, respectively), which we attribute primarily to boundary errors introduced during field detection. Notably, the precision for EIP and HTTP is close to $1.00$, indicating that when field boundaries are correctly identified, the inferred function labels are highly reliable. S7Comm is particularly challenging because messages over TCP/IP are encapsulated by multiple headers, creating a deeply nested structure in which semantic errors at outer layers can propagate to inner sub-payloads.

\begin{answerbox}
    \textit{Answer to RQ2:} \codename identifies both field type and function with high accuracy, significantly outperforming all the state-of-the-art PRE techniques.
\end{answerbox}

\subsection{RQ3: Robustness and Generalization}

\codename uses GPT-o3 model as its backbone LLM, which raises the concern that the model may have been exposed to protocol specifications during pre-training, potentially introducing bias into our evaluation. To decouple \codename's performance from any latent prior knowledge, we construct a \emph{zero-prior-knowledge} setting. Concretely, our prompt explicitly forbids the use of any background knowledge about Modbus, S7Comm, Ethernet/IP, DNP3, DNS, FTP, TFTP, and HTTP, and requires field extraction to be derived solely from the provided message content and auxiliary traces. We validate that this constraint is effective by observing a clear performance degradation when running vanilla GPT-o3 under the zero-knowledge prompt (Table~\ref{tab:naive_llm_perf}). We then evaluate three configurations on the D1 dataset: \codename, \codename-zero, and BinPRE. As shown in Table~\ref{tab:sempre_wo_bg}, \codename indeed benefits from background knowledge. However, even without it, \codename-zero still significantly outperforms BinPRE on most protocols. The only exception is DNP3, where API input-length limits force us to truncate the taint context, reducing the amount of usable dynamic evidence and thereby weakening the method's effectiveness.
\begin{table}[H]
    \centering
    \caption{Effectiveness w/wo Prior Knowledge.}\label{tab:sempre_wo_bg}
    \addtolength{\tabcolsep}{-2pt}
    \resizebox{\linewidth}{!}{
    \begin{tabular}{rccc|ccc|ccc}
    \toprule
        \multirow{2}{*}{Protocols} &\multicolumn{3}{c}{\codename} & \multicolumn{3}{c}{$\text{\codename-{zero}}^1$} & \multicolumn{3}{c}{BinPRE} \\\cmidrule{2-10}
        & Field & Type & Fn. & Fields & Type & Fn. & Fields & Type & Fn.   \\ \midrule
        Modbus & 0.99 & \textbf{0.81} & \textbf{1.00} & 0.99 & 0.73 & 0.99 & \textbf{1.00} & 0.41 & \textbf{1.00}  \\ 
        S7comm & 0.81 & \textbf{0.80} & 0.64 & \textbf{0.82} & 0.59 & 0.35 & 0.05 & 0.00 & 0.00  \\ 
        Ethernet/IP & \textbf{0.80} & \textbf{0.69} & \textbf{1.00} & 0.62 & 0.62 & 0.99 & 0.25 & 0.22 & 1.00  \\ 
        DNP3 & \textbf{0.80} & 0.82 & 0.90 & 0.64 & 0.34 & 0.31 & 0.70 & 0.00 & 0.86  \\ 
        DNS & \textbf{0.72} & \textbf{0.71 }& \textbf{0.93} & 0.65 & 0.55 & 0.72 & 0.56 & 0.49 & 0.75  \\ 
        FTP & \textbf{0.66} & \textbf{0.72} & \textbf{0.77} & 0.63 & 0.68 & 0.74 & 0.37 & 0.00 & 0.00  \\ 
        TFTP & \textbf{1.00} & 0.35 & \textbf{1.00} & \textbf{1.00} & \textbf{0.41} & \textbf{1.00} & 0.54 & 0.27 & 0.64  \\ 
        HTTP & 0.60 & \textbf{0.62} & 0.70 & \textbf{0.85} & 0.55 & 0.54 & 0.31 & 0.00 & 0.00  \\ \midrule
        Average & \textbf{0.76} & \textbf{0.72} & \textbf{0.74} & 0.72 & 0.45 & 0.43 & 0.39 & 0.14 & 0.35 \\ \bottomrule
    \end{tabular}
    }
    \flushleft \scriptsize 1 - \codename-zero: \codename in the zero-prior-knowledge setting.
\end{table}

We further evaluate on undocumented protocols (D2 dataset), whose specifications are unlikely to be memorized by LLMs. As shown in Table~\ref{tab:undoc_perf} in Appendix, \codename consistently outperforms both BinPRE and vanilla GPT-o3. Although \codename and BinPRE rely on the same taint information, BinPRE groups bytes using the Needleman–Wunsch algorithm, which lacks instruction-level semantic awareness and often produces incorrect merges. In contrast, \codename leverages LLM-guided consistency between field detection and semantic inference to interpret taint outputs and delineate boundaries more accurately. An exception applies to KTP, where performance is comparable between \codename and vanilla GPT-o3, because half of the fields are firmly associated with the sending end, while the other fields work for the receiving end, and taint information affects only fields related to the sending end.

Finally, we replace GPT-o3 with DeepSeek-v3.1-terminus to examine whether \codename's effectiveness changes substantially under a different backbone LLM. The results remain comparable and consistently outperform BinPRE (Table~\ref{tab:backbones_llm_perf}), suggesting that \codename's gains are not tied to any single model, but stem primarily from its overall design.

\begin{table}[!ht]
    \centering
    \caption{GPT-o3 Performance.}
    \label{tab:naive_llm_perf}
    \addtolength{\tabcolsep}{-3pt}
    \resizebox{0.9\linewidth}{!}{
    \begin{tabular}{rccc|ccc}
    \toprule
        \multirow{2}{*}{Protocols} & \multicolumn{3}{c}{w Knowledge}   & \multicolumn{3}{c}{w/o Knowledge} \\ \cmidrule{2-7} 
         & Fields  & Types  & Fn.  & Fields  & Types  & Fn.   \\ \midrule 
        Modbus & \textbf{0.67} & \textbf{0.63} & 0.45 & 0.62 & 0.62 & \textbf{0.57} \\ 
        S7comm & \textbf{0.80} & \textbf{0.77} & \textbf{0.40} & 0.18 & 0.19 & 0.06  \\
        Ethernet/IP & \textbf{0.38} & \textbf{0.38} & \textbf{0.12} & 0.32 & 0.23 & 0.06 \\ 
        DNP3 & 0.27 & 0.36 & 0.27 & \textbf{0.36 }& \textbf{0.54} & \textbf{0.33} \\ 
        DNS & 0.70 & 0.84 & 0.39 & \textbf{0.87} &\textbf{ 0.89} & \textbf{0.51}  \\ 
        FTP & \textbf{0.48} & \textbf{0.79} & \textbf{0.85} & 0.38 & 0.67 & 0.74\\ 
        TFTP & \textbf{1.00} & \textbf{0.69} & \textbf{1.00 }& 0.77 & 0.62 & 0.91   \\ 
        HTTP  & 0.48 & 0.64 & 0.21 & \textbf{0.50} & \textbf{0.66} & \textbf{0.24}   \\ \midrule
        Average & \textbf{0.53} & \textbf{0.58} & \textbf{0.34} & 0.44 & 0.46 & 0.29  \\ \bottomrule
    \end{tabular}}
\end{table}


\begin{answerbox}
    \textit{Answer to RQ3:} \codename maintains strong performance without prior LLM knowledge or with a different backbone LLM, consistently outperforming SOTA techniques.
\end{answerbox}

\subsection{RQ4: Ablation Study}
In this study, we explore the effectiveness of two core components of \codename: recursive sub-payload detection (Figure~\ref{fig:field}) and iterative optimization (Figure~\ref{fig:semantic}). To conduct this ablation study cost-efficiently, we use a small-scale dataset derived from RQ1 by selecting the first 10 messages from each protocol. We evaluate the following configurations of \codename:
\begin{itemize}[leftmargin=*]
    \item \textbf{\codename-norec-noop}: remove both the recursive sub-payload detection and the iterative optimization
    \item \textbf{\codename-noop}: remove the iterative optimization
    \item \textbf{\codename-ToT}: remove the ToT
    \item \textbf{\codename-norec}: remove the recursive sub-payload detection
    \item \textbf{\codename}: full-fledged \codename
\end{itemize}

Recursive sub-payload detection addresses the challenge of field detection in the presence of nested payloads. As shown in Table~\ref{tab:ablation_field} in Appendix, adding this component increases the average F1 score from 0.58 to 0.76 compared to \codename-norec-noop. Although DNS sees a decrease (1.00 to 0.77), substantial gains are achieved for Ethernet/IP ($0.37 \to 0.71$) and HTTP ($0.39 \to 0.62$). A similar performance shift between \codename-norec-noop and \codename further confirms the effectiveness of recursive detection.


\begin{table}[h]\centering
\caption{Ablation Study Results - Semantic Inference.}
\label{tab:ablation_type_fn}
\addtolength{\tabcolsep}{-3pt}
\resizebox{\linewidth}{!}{
\begin{tabular}{rcc|cc|cc}\toprule
\multirow{2}{*}{Protocols}  &\multicolumn{2}{c}{\textbf{\codename-norec-noop}} &\multicolumn{2}{c}{\textbf{\codename-noop}} &\multicolumn{2}{c}{\textbf{\codename}} \\\cmidrule{2-7}
&Type F1 & Fn. F1 &Type F1 & Fn. F1 &Type F1 & Fn. F1 \\\midrule
Modbus  &\textbf{0.79 } &\textbf{1.00} &0.78 &\textbf{1.00} &0.78 &\textbf{1.00} \\
S7comm  &0.68 &0.53 &0.86 &0.70 &\textbf{0.91}  &\textbf{0.77} \\
Ethernet/IP &0.28 &0.25 &0.31 &0.24  &\textbf{0.47} &\textbf{0.62} \\
DNP3  &0.75 &0.82 &\textbf{0.84} &\textbf{0.97} &0.80 &0.90 \\
DNS &0.71 &0.87  &\textbf{0.72} &\textbf{1.00} &0.70 &0.96 \\
FTP  &0.68 &0.71  &\textbf{0.68} &\textbf{0.71} &0.67 &\textbf{0.71} \\
TFTP &0.50 &0.97 &\textbf{0.40} &0.95 &\textbf{0.40} &\textbf{1.00} \\
HTTP &0.31 &0.18 &0.65  &0.49  &\textbf{0.68} &\textbf{0.71} \\
\midrule
Average  &0.48 &0.51 &0.56 &0.58  &\textbf{0.62} &\textbf{0.74} \\
\bottomrule
\end{tabular}}
\end{table}

\begin{table}[h]\centering
\caption{Ablation Study Results - Field Detection.}\label{tab:ablation_field}
\addtolength{\tabcolsep}{-3pt}
\resizebox{\linewidth}{!}{
\begin{tabular}{rrrc|rrr|rrr}\toprule
\multirow{2}{*}{Protocols}  &\multicolumn{3}{c}{\textbf{\codename-norec-noop}} &\multicolumn{3}{c}{\textbf{\codename-noop}} &\multicolumn{3}{c}{\textbf{\codename}} \\\cmidrule{2-10}
&Pre. &Rec. &F1 &Pre. &Rec. &F1 &Pre. &Rec. &F1 \\\midrule
Modbus &1.00 &1.00 &\textbf{1.00} &1.00 &1.00 &\textbf{1.00} &1.00 &0.84 &0.91 \\
S7comm &0.98 &0.69 &0.81 &0.90 &0.95 &\textbf{0.93} &0.89 &0.94 &0.92 \\
Ethernet/IP &0.96 &0.23 &0.37 &0.81 &0.63 &0.71 &0.87 &0.61 &\textbf{0.72} \\
DNP3 &0.95 &0.58 &0.72 &0.91 &0.94 &\textbf{0.92} &0.90 &0.78 &0.84 \\
DNS &1.00 &1.00 &1.00 &0.63 &1.00 &\textbf{0.77} &0.58 &1.00 &0.73 \\
FTP &0.34 &0.78 &0.47 &0.34 &0.78 &\textbf{0.47} &0.23 &0.53 &0.32 \\
TFTP &1.00 &1.00 &1.00 &0.91 &1.00 &0.95 &1.00 &1.00 &\textbf{1.00} \\
HTTP &0.53 &0.31 &0.39 &0.61 &0.62 &0.62 &0.81 &0.68 &\textbf{0.74} \\
\midrule
Average &0.84 &0.45 &0.58 &0.77 &0.75 &\textbf{0.76} &0.81 &0.72 &\textbf{0.76} \\
\bottomrule
\end{tabular}}
\end{table}

Improved field detection should also enhance semantic inference, as intended. \codename-noop outperforms \codename-norec-noop on semantic tasks, raising average F1 $0.48 \to 0.56,$ for type inference and $ 0.51 \to 0.58$ for function inference (Table~\ref{tab:ablation_type_fn}). Adding iterative optimization further boosts performance,  raising F1 scores to 0.62 for type inference and 0.74 for function inference. While some protocols show minor drops, the gains on Ethernet/IP and HTTP are substantial, as shown in Table~\ref{tab:ablation_type_fn}.

From these results, we observe that recursive sub-payload detection is effective: the total correct fields are 176 without sub-payload detection (vs. 483 in SemPRE) for the nested protocol Ethernet/IP. To strengthen the component's contribution, we build a small nested dataset $D4$ and evaluate \codename-norec and \codename on that. The results are shown in Table~\ref{tab:ablation_rec}.

\begin{table}[]
\caption{Ablation Study Results - Sub-payload Detection.}\label{tab:ablation_rec}
\resizebox{\linewidth}{!}{
\begin{tabular}{cccc|ccc|ccc}
\toprule
\multirow{2}{*}{Protocols} & \multicolumn{3}{c}{SemPRE-norec} & \multicolumn{3}{c}{SemPRE} & \multicolumn{3}{c}{BinPRE}\\ \cmidrule{2-10}
 & bnd F1 & type F1 & fn F1 & bnd F1 & type F1 & fn F1 & bnd F1 & type F1 & fn F1 \\ \midrule
Ethernet/IP     & 0.55 & 0.12 & 0.13 & \textbf{0.72} & \textbf{0.47} & \textbf{0.62} &0.19 & 0.00 & 0.13 \\

BAC   & \textbf{0.92 } & \textbf{0.80} & \textbf{0.96} & 0.84 & \textbf{0.80} & 0.90 & 0.79 & 0.21 & 0.15 \\
OPC    & \textbf{0.78} & \textbf{0.72} & {0.95} & 0.73                           & 0.70                            & \textbf{0.96} & 0.36 & 0.00 & 0.00 \\

\midrule
Avg.   & 0.71                           & 0.50                            & 0.56                          &\textbf{0.76}                           & \textbf{0.62 }                           & \textbf{0.74} &0.29 &0.06 & 0.12 \\ 
\bottomrule
\end{tabular}}
\end{table}

\begin{table}[]
\caption{Ablation Study Results - ToT Affect.}
\label{tab:ablation_tot}
\resizebox{\linewidth}{!}{
\begin{tabular}{cccc|ccc}
\toprule
\multirow{2}{*}{Protocols} & \multicolumn{3}{c}{SemPRE-ToT} & \multicolumn{3}{c}{SemPRE}\\ \cmidrule{2-7}
 & bnd F1 & type F1 & fn F1 & bnd F1 & type F1 & fn F1 \\ \midrule
modbus & \textbf{1.00} & 0.78 & \textbf{1.00 }& 0.91 & 0.78 & \textbf{1.00} \\
s7     & 0.89 & 0.85 & 0.71 &\textbf{0.92}& \textbf{0.91} & \textbf{0.77 } \\
eip    & 0.55 & 0.12 & 0.13 & \textbf{0.72} & \textbf{0.47} & \textbf{0.62} \\

dnp3   & \textbf{0.92 } & \textbf{0.80} & \textbf{0.96} & 0.84 & \textbf{0.80} & 0.90 \\
dns    & \textbf{0.78} & \textbf{0.72} & {0.95} & 0.73                           & 0.70                            & \textbf{0.96}                          \\

ftp    & \textbf{0.45}                           & 0.61                            & 0.67                          & 0.32                           &\textbf{ 0.67}                            &\textbf{0.71}                          \\
 
tftp   & \textbf{1.00} & \textbf{0.40} & \textbf{1.00 }                         & \textbf{1.00}                           & \textbf{0.40}                            & \textbf{1.00}                          \\

http   & 0.72                           & 0.66                            & 0.62                          & \textbf{0.74}                           & \textbf{0.68}                            & \textbf{0.71}                          \\
\midrule
Avg.   & 0.71                           & 0.50                            & 0.56                          &\textbf{0.76}                           & \textbf{0.62 }                           & \textbf{0.74} \\ 
\bottomrule
\end{tabular}}
\end{table}

\codename-ToT degrades F1 from 0.76/0.62/0.74 to 0.71/0.50/0.56 (Table~\ref{tab:ablation_tot}, compared to \codename, indicating that ToT alone cannot improve performance but plays a key role in \codename.

Overall, these results demonstrate that both recursive sub-payload detection and iterative optimization significantly enhance \codename’s ability to extract fields and infer semantics.

\begin{answerbox}
    \textit{Answer to RQ4:} \codename’s recursive sub-payload detection and iterative optimization are essential for high performance.
\end{answerbox}

\subsection{RQ5: Case Study}
Finally, we evaluate \codename's efficacy using a case study on a variant of the Mirai botnet~\cite{mirai-source-code}, a well-known malware family that targets Internet-of-Things (IoT) devices and poses a persistent threat in the cybersecurity landscape. We simulate a realistic analysis scenario in which only the Mirai binary and captured network traffic are available. We analyze 10 messages exchanged between the command-and-control (C\&C) server and an infected device, reflecting an early-stage reverse-engineering setting in which no protocol documentation is assumed. We evaluate both \codename and BinPRE against a manually constructed ground truth derived from the source code. Across the ten messages, the ground truth contains 117 fields, 127 type instances, and 45 function labels. 


BinPRE assumes that a program consumes each message buffer in a single pass. The Mirai binary, however, reuses the same buffer and parses it a second time, breaking BinPRE's offset arithmetic and leading to almost zero accuracy. In contrast, although \codename and BinPRE rely on the same taint outputs, \codename does not impose a single-pass assumption. Instead, it uses LLM-based reasoning over the collected evidence and can therefore accommodate repeated buffer accesses. To isolate this effect, we run an additional experiment in which both tools are explicitly informed that the Mirai binary traverses the message buffer twice. With this hint, BinPRE achieves an F1 score of $0.55$ for field detection. By comparison, \codename attains a stable $0.83$ F1 for field detection with or without the hint. At the per-message level, \codename recovers perfectly accurate boundaries for 9 out of 10 messages, whereas BinPRE does so for only 2. The remaining gap largely stems from BinPRE's dependence on the Needleman--Wunsch (NW) algorithm for aligning instruction sequences, as it captures primarily instruction-level correspondences and often misses deeper semantic relations needed to correctly merge and split tainted regions.

\begin{table}[!htp]\centering
\caption{Case Study - F1 Scores.}\label{tab:case_f1}
\resizebox{0.5\linewidth}{!}{
\begin{tabular}{r|c|c|c}\toprule
& Field & Type & Function \\
\midrule
\codename &\textbf{0.83}&\textbf{0.69}&\textbf{0.74} \\
BinPRE &0.55&0.27&0.19 \\
\bottomrule
\end{tabular}}
\end{table}

\begin{answerbox}
    \textit{Answer to RQ5:} The \codename can effectively support real-world security analysis, particularly in scenarios where protocols are undocumented and malicious.
\end{answerbox}
\section{Related Work}

\subsection{Protocol Reverse Engineering} 
Automated protocol reverse engineering (PRE) techniques broadly fall into two categories: \emph{network-traffic-based} and \emph{binary-analysis-based} approaches. Traffic-based systems infer message layouts by identifying recurring byte patterns across packet traces, whereas binary-analysis methods reconstruct formats and semantics by examining how a program processes its inputs. Early binary-analysis tools (e.g., Polyglot~\cite{caballero2007polyglot}, AutoFormat~\cite{lin2008automatic}, and Tupni~\cite{cui2008tupni}) use taint tracking and execution context analysis to extract field boundaries but provide limited semantic information. In parallel, traffic-based methods have evolved from largely heuristic systems such as NEMESYS~\cite{kleber2018nemesys} to more structured approaches like NetPlier~\cite{ye2021netplier}, which leverage sequence alignment and probabilistic inference to identify type-indicating fields and reconstruct protocol state machines. Recent work has improved both semantic depth and robustness. BinaryInferno~\cite{chandler2023binaryinferno} combines multiple heuristics in an entropy-guided ensemble to improve precision and completeness, while DynPRE~\cite{luo2024dynpre} adopts an active-learning strategy that generates probe messages to explore protocol states and reveal richer semantics. BinPRE~\cite{jiang2024binpre} represents the state of the art in binary-analysis-based PRE. It aligns instruction sequences, employs atomic field detectors for common data-handling idioms, and iteratively clusters and refines candidate fields. Evaluated on eight real-world protocols, BinPRE surpasses prior systems, including Polyglot, AutoFormat, Tupni, BinaryInferno, and DynPRE, in both precision and recall.
\subsection{LLM Reasoning Techniques}
Early progress in LLM reasoning stemmed solely from prompt engineering. Chain-of-Thought~\cite{wei2022chain} (CoT) prompting revealed that inserting a few worked examples elicits multi-step reasoning, boosting accuracy on math and commonsense tasks. Self-Consistency~\cite{wang2022self} improves CoT by sampling diverse reasoning paths and selecting the majority answer, yielding double-digit gains across several benchmarks. Active-Prompt~\cite{diao2023active} further adapts the exemplar set on-the-fly, choosing those that best suit a new task. Despite these advances, all three methods rely on a single linear trace of text tokens and quickly degrade when the input interleaves structured bytes with noisy instrumentation—precisely the scenario in protocol reverse engineering (PRE).

To avoid the myopia of linear CoT, Tree-of-Thoughts~\cite{yao2023tree} (ToT) treats reasoning as an explicit tree search, allowing the model to branch, backtrack, and evaluate partial solutions before committing; GPT-4 accuracy on the “24 Game” rises from 4\% to 74\% under this scheme. Yet ToT still assumes that each “thought” is a clean natural-language fragment, leaving open how to segment—and reason over—raw byte streams mixed with taint-trace noise. \codename tackles exactly this gap by anchoring tree exploration in execution evidence.

LLMs can also grade and revise their own work. A recent study~\cite{renze2024self} shows that introducing a reflection loop—generate, critique, then refine—mainly lifts performance on multiple-choice QA and planning benchmarks. \codename borrows this idea but grounds the critique in concrete taint maps and loop inventories, rather than in free-form textual judgment.

\section{Discussion}

\vspace{4pt}\noindent \textbf{LLM Limitations.}
Despite their impressive performance across a wide range of tasks, large language models (LLMs) are subject to several important limitations, particularly in terms of reliability and factual accuracy. One of the most critical issues is hallucination, where the model generates fluent but factually incorrect or fabricated outputs. Studies have shown that LLMs may confidently generate incorrect or hallucinatory answers~\cite{slobodkin2023curious, xiong2024can}, produce inaccurate or fabricated sources~\cite{walters2023fabrication}, and exhibit reduced factual reliability when the required knowledge is insufficiently represented in their pre-training data~\cite{kandpal2023large, lin2022truthfulqa}. Furthermore, LLMs often lack grounding in external knowledge and can struggle with consistency across long or complex reasoning chains~\cite{achiam2023gpt}. Although we apply Tree-of-Thoughts and iterative refinement to improve the LLM's effectiveness, the LLM's limitations are hard to eliminate entirely. 

\vspace{4pt}\noindent \textbf{Byte-Level v.s Bit-Level Semantics.} 
Our current implementation supports only six semantic function labels for fair comparison with BinPRE, far fewer than those found in real-world protocols. In addition, \codename performs semantic annotation at \emph{byte} granularity. This design follows directly from our reuse of BinPRE's taint-analysis backend, which tracks data dependencies at the byte level. While byte-level tainting integrates cleanly with dynamic program analysis and remains efficient in practice, it limits \codename's ability to recover finer-grained semantics. This limitation is particularly salient for embedded and low-level protocols, where meaning is often encoded at the \emph{bit} level rather than the \emph {byte} level. For example, in the Controller Area Network (CAN) protocol, individual bits within a byte frequently represent distinct flags or sensor states (e.g., fault indicators and enable/disable toggles) \cite{bosch1991can}. Such semantics cannot be fully captured with byte-level inference alone, which constrains \codename's precision in these settings. Extending \codename to support bit-level taint tracking, bit-level semantic inference, and a richer taxonomy of semantic functions is an important direction for future work.

\vspace{4pt}\noindent \textbf{Robustness to Adversarial Settings.} 
Obfuscation techniques that randomize field access order or copy data into temporary scratch buffers can blur taint signals and defeat ToT search. This problem involves defeating obfuscation, which is fundamentally orthogonal to the protocol reverse engineering task. We will leave this for future research. 

\section{Conclusion}

Protocol reverse engineering (PRE) reconstructs message structures and field semantics, yet existing tools rely on fragile heuristics and struggle with nested payloads and semantic dependencies. We present \codename, a fully automated PRE framework that couples fine-grained dynamic analysis with a recursive LLM reasoning loop. Using taint-flow and loop traces, it segments messages, detects nested payloads, and iteratively refines structure and semantics through a monotonic EM-style process. On eight mainstream protocols, \codename attains an F1 score of 0.76 for field detection—over twice the prior best—and raises semantic-inference accuracy to 0.72. It also proves effective on Mirai botnet traffic, accurately recovering undocumented fields and semantics. These results show that LLM-based reasoning can replace brittle heuristics and set a new state of the art in PRE.





\cleardoublepage
\appendix
\section*{Ethical Considerations}

\noindent \textbf{Stakeholder Identification and Impacts.}
The research process and outcomes affect multiple stakeholders:

\vspace{4pt}\noindent \textit{Research Team.} Benefits from academic advancement and technical validation; bears responsibility for guiding ethical use.

\vspace{4pt}\noindent \textit{Security Researchers/Analysts.} Gains a high-precision tool for vulnerability discovery, malware analysis, and IDS signature generation, improving work efficiency.

\vspace{4pt}\noindent \textit{Enterprises/Organizations.} Can leverage \codename to identify protocol vulnerabilities, but faces potential risks of legitimate protocol reverse-engineering by third parties.

\vspace{4pt}\noindent \textit{General Users.} Indirectly benefits from enhanced system security but may face privacy/property risks if the tool is misused.

\vspace{4pt}\noindent \textit{Open Source Community.} Obtains open-source resources to advance PRE research, promoting field development.

\vspace{4pt}\noindent \textbf{Security Community and Society}

\vspace{4pt}\noindent \textit{Beneficence.} \codename addresses limitations of heuristic-based PRE tools, doubling field detection F1 score (0.76 vs. <0.40) and improving semantic inference accuracy (0.72 vs. 0.14), advancing network security defense capabilities.

\vspace{4pt}\noindent \textit{Respect for Persons.} All experimental datasets are from open-source repositories or legally captured traffic, with sensitive information (IP addresses, device IDs) anonymized to protect user privacy.

\vspace{4pt}\noindent \textit{Justice.} The \codename prototype is open-sourced to ensure fair access to technical resources, with transparent experimental design for community verification.

\vspace{4pt}\noindent \textit{Respect for Law and Public Interest.} The research and publication comply with network security regulations; we explicitly oppose misuse of the tool for illegal activities.

\vspace{4pt}\noindent \textbf{Potential Harms Analysis.}
We acknowledge that the protocols may create negative impacts if used by illicit actors to reveal private protocols. For example, an attacker may reconstruct the syntactic and semantic information of a private protocol using \codename and then send spurious messages to attack normal users.

\vspace{4pt}\noindent \textbf{Ethical Rationale for Research and Publication}

\vspace{4pt}\noindent \textit{Research Justification.} Traditional PRE tools rely on brittle heuristics, failing to handle nested protocols and semantic dependencies. \codename fills this technical gap with LLM-driven joint optimization, critical for improving security defense.

\vspace{4pt}\noindent \textit{Publication Justification.} Open-sourcing and publishing enable academic exchange, allowing the community to refine the tool. Ethical guidelines and positive application scenario descriptions minimize misuse risks, ensuring alignment with public interest.


\section*{Open Science}
Our code is available at https://anonymous.4open.science/r/SemPRE-F9F8.
\cleardoublepage

\bibliographystyle{IEEEtran}
\bibliography{ref}

\appendix
\cleardoublepage
\section{Appendices}
\subsection{Prompt Design}
\label{sec:prompt_tmp}
\noindent \textbf{Prompting.} 
We use structured prompt templates to guide the LLM through field detection, sub-payload identification, and semantic inference. Beyond a fixed introduction, each prompt includes (i) taint and loop analysis outputs, (ii) the raw message, (iii) high-confidence semantic cues, and (iv) explicit reasoning instructions. This structure constrains the model to task-specific context, reducing hallucinations and improving consistency across runs. At initialization, we instruct the LLM to act as an expert in protocol reverse engineering.

\subsubsection{The Prompt Template: Field Detection }
\begin{figure}[!h]
    \centering
    \includegraphics[width=0.95\linewidth]{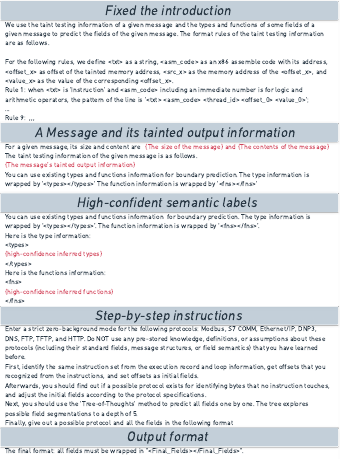}
    \caption{Prompt Structure: Field Detection Template}
    \label{fig:field_prompt}
\end{figure}

The prompt template for the task of field detection comprises by five parts, shown in Figure~\ref{fig:field_prompt}, which are titled in italic font. We set Tree-of-Thoughts (ToT) parameters in the `step-by-step instructions' section to guide LLM reasoning. 

\subsubsection{The Conversation for Field Detection}
\begin{figure}[h]
    \centering
    \includegraphics[width=0.95\linewidth]{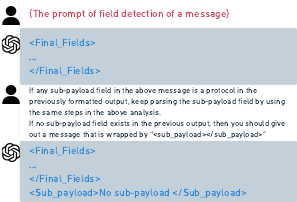}
    \caption{LLM Interaction: Field Detection Process}
    \label{fig:sub_prompt}
\end{figure}

We start a conversation with the LLM for field detection, shown in Figure~\ref{fig:sub_prompt}. With the context, the LLM only reasons the untainted bytes.

\subsubsection{The Prompt Template: Semantic Inference}
\begin{figure}[h]
    \centering
    \includegraphics[width=0.83\linewidth]{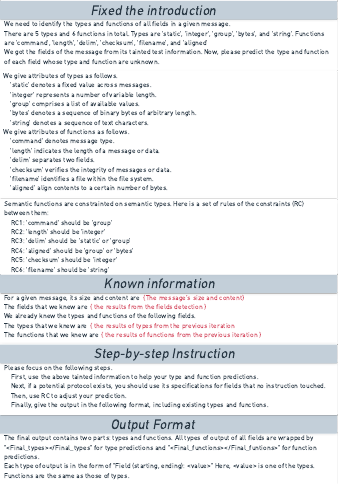}
    \caption{Prompt Structure: Semantic Inference Template}
    \label{fig:sem_prompt}
\end{figure}

The template for semantic inference comprises by four parts, as shown in Figure~\ref{fig:sem_prompt}, which are titled in italic font. `known information' includes a message, its fields, high-confident semantic labels.

\subsubsection{The Conversation for Semantic Inference}
\begin{figure}[h]
    \centering
    \includegraphics[width=\linewidth]{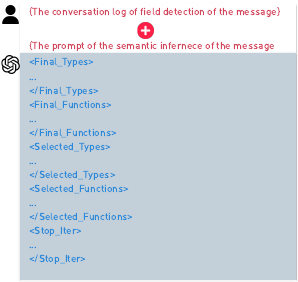}
    \caption{LLM Interaction: Semantic Inference Process}
    \label{fig:sem_infer}
\end{figure}

In each iteration, we apply the field-detection and semantic-inference prompts to obtain updated fields and semantic labels. The taint information continues to guide the LLM by enforcing consistency between the detected fields and their inferred semantics. The structure is in Figure~\ref{fig:sem_infer}


\subsection{Tables}
\subsection{Common Types and Functions}
We follow the definitions of types and functions as identified in prior work~\cite{jiang2024binpre}. Details of definitions are in Table~\ref{tab:semantic-types-functions}
\begin{table}[!ht]
\centering
\caption{Semantic types and functions}
\label{tab:semantic-types-functions}
\begin{tabular}{l|p{0.75\linewidth}}
\hline
\multicolumn{2}{c}{\textbf{Semantic Types}} \\ \hline
\textit{Static}   & Fixed values across messages, usually protocol-specific flags \\ \hline
\textit{Integer}  & Numeric metadata (e.g., sizes), in either big- or little-endian \\ \hline
\textit{Group}    & Enumerated static values for actions, message types, etc. \\ \hline
\textit{Bytes}    & Protocol data in raw binary format \\ \hline
\textit{String}   & Text character sequences (e.g., file names) \\ \hline
\multicolumn{2}{c}{\textbf{Semantic Functions}} \\ \hline
\textit{Command}  & Denotes message types \\ \hline
\textit{Length}   & Indicates the length of messages or data \\ \hline
\textit{Delim}    & Separates two fields \\ \hline
\textit{Checksum} & Verifies message or data integrity \\ \hline
\textit{Filename} & Identifies a file within the file system \\ \hline
\textit{Aligned}  & Aligns contents to a specified byte boundary \\ \hline
\end{tabular}
\end{table}

\subsubsection{Results from Different Tools}
We evaluate the examples from the section~\ref{sec:motivation} using multiple tools and summarize the results in Tables~\ref{tab:cmp_fname_tftp}--\ref{tab:cmp_len_s7}--\ref{tab:cmp_cip_eip}. We compare \codename against BinPRE, Polyglot, and Tupni (BPT) to highlight both the strengths of \codename and the limitations of BPT. As shown in Table~\ref{tab:cmp_fname_tftp}, the results illustrate the shortcomings of BPT’s inflexible, heuristic-based rules.

\begin{table}[!ht]
    \centering
    \caption{Different Results on the Filename Field of TFTP from Different Tools}
   {
    \begin{tabular}{ccc}\toprule
         Tools& Field & Function \\ \midrule
         BinPRE &$[2], [3,17]$ & None, Aligned\\
         Polyglot & $[2],[3],[4],...,[16],[17]$  & -\\
         Tupni & $[2], [3,17]$ & -\\ 
         \codename & $[2, 17]$& Filename \\ \midrule
         GroundTruth & $[2, 17]$& Filename\\ 
    \end{tabular}
    }
    \label{tab:cmp_fname_tftp}
\end{table}

Next, we observe the effects of the ignorance of semantic relationships in BPT. However, \codename can correctly rebuild them in Table~\ref{tab:cmp_len_s7}.
\begin{table}[!ht]
    \centering
    \caption{Different Results on Two Length Fields of S7 from Different Tools}
    \begin{tabular}{ccc}\toprule
         Tools& Field & Function \\ \midrule
         BinPRE & $\triangle$ & None, None \\
         Polyglot & $\triangle$ & None, None \\
         Tupni & $\triangle$ & None, None \\ 
         \codename & $[2, 3], [4]$ & Length, Length\\ \midrule
         GroundTruth & $[2, 3], [4]$& Length, Length \\
    \end{tabular}
    \label{tab:cmp_len_s7}
    \flushleft
    $\triangle$: incorrect field detection; None: the lack of prediction
\end{table}

Finally, we spot the power of \codename about detecting sub-payload but BPT can't in Table~\ref{tab:cmp_cip_eip}
\begin{table}[!ht]
    \centering
    \caption{Different Results on Nested Protocol of EIP from Different Tools}
    \begin{tabular}{ccc}\toprule
         Tools& Field & Nested Protocol \\ \midrule
         BinPRE & $[30, 81] $ & - \\
         Polyglot & $\triangle$ & - \\
         Tupni & $\triangle$ & - \\ 
         \codename & $[46, 81]$ & D. \\ \midrule
         GroundTruth & $[46, 81]$& CIP \\
    \end{tabular}
    \label{tab:cmp_cip_eip}
    \flushleft
    $\triangle$: incorrect field detection: -: not support; D.: detect a nested sub-payload; 
\end{table}

\subsubsection{Details for Protocol in Datasets}
Table~\ref{tab:messages} shows the details of the protocols in the datasets. Format colomn indicates the categories of the content of procotol messages: binary, text, or mixed. Project colomn shows the resources of the protocol binary file.
\begin{table}[!ht]
    \centering
    \caption{Protocol Datasets for Evaluation.}
    \label{tab:messages}
    \addtolength{\tabcolsep}{-4pt}
    \resizebox{\linewidth}{!}{
    \begin{tabular}{ccccc}
        \toprule
        Dataset & Protocol & Format & Project & Message Source \\
        \midrule
        \multirow{8}{*}{D1} & Modbus & binary & \texttt{freemodbus}~\cite{freemodbus} & Open-source~\cite{modbus-dns-trace} \\
        &S7comm & binary & \texttt{snap7}~\cite{snap7} & Open-source~\cite{s7-trace} \\
        &Ethernet/IP & binary & \texttt{OpENer}~\cite{opener} & Open-source~\cite{eip-trace} \\
        &DNP3.0 & binary & \texttt{opendnp3}~\cite{opendnp3} & \texttt{tcpdump} \\
        &FTP & text & \texttt{LightFTP}~\cite{lightftp} & Open-source~\cite{ftp-trace} \\
        &TFTP & mixed & \texttt{tftpd-hpa}~\cite{tftp-hpa} & Open-source~\cite{tftp-trace} \\
        &DNS & mixed & \texttt{dnsmasq}~\cite{dnsmasq} & Open-source~\cite{modbus-dns-trace} \\
        &HTTP & text & \texttt{miniweb}~\cite{miniweb} & \texttt{tcpdump} \\ \midrule
        \multirow{3}{*}{D2}&KTP & binary & \texttt{KTP-Source-Code}~\cite{ktp-source-code} & \texttt{tcpdump} \\
        &CCN & binary & \texttt{CCN-Source-Code}~\cite{ccn-source-code} & \texttt{tcpdump} \\
        &Xfer & mixed & \texttt{Xfer-Source-Code}~\cite{xfer-source-code} & \texttt{tcpdump} \\ \midrule
        \multirow{1}{*}{D3}&Mirai & mixed & \texttt{Mirai-Source-Code}~\cite{mirai-source-code} & \texttt{tcpdump} \\
        \bottomrule
    \end{tabular}}
\end{table}

\subsubsection{Semantic Capabilities of the Baselines}
We summarize the semantic modeling capabilities of representative prior protocol reverse engineering (PRE) systems. We compare these approaches along three dimensions: (1) the analysis paradigm they adopt, (2) the kinds of semantic types for protocol fields they infer, and (3) the kinds of semantic functions they are able to identify. Details are in Table~\ref{tab:cap_baselines}.

\begin{table}[!ht]
    \centering
    \caption{Capability of baselines and \codename}
    \label{tab:cap_baselines}
    \resizebox{\linewidth}{!}{
    \begin{tabular}{llcl}
    \hline
        Protocols & Approaches & Types & Functions \\ \midrule
        PolyGlot & binary-analysis-based  & - & CMD; LEN; DELIM \\ 
        Tupni & binary-analysis-based  & - & LEN; CHS \\ 
        AutoFormat & binary-analysis-based  & - & - \\ 
        BinPRE & binary-analysis-based  & $COMM_{Types}$ & $COMM_{Fns}$ \\
        \codename & binary-analysis-based  & $COMM_{Types}$ & $COMM_{Fns}$ \\ \midrule
        BinaryInferno & network-traffic-based & - & LEN \\ 
        DynPRE  & network-traffic-based & - & CMD, LEN \\ 
    \end{tabular}
    }
    \flushleft
    \scriptsize
    -: not support; CMD: command; LEN: length; DELIM: delimitor; CHS: checksum; $COMM\_{Types}$: common types; $COMM\_{Fns}$: common functions.
\end{table}


\subsubsection{Tables Related to RQ3}
Table~\ref{tab:backbones_llm_perf} shows the results for \codename with different backbones. 
Results show that the performance of \codename does not depend on a specific LLM model. 
\begin{table}[!htp]
    \centering
    \caption{Performance with Different Backbones.}
    \label{tab:backbones_llm_perf}
    \addtolength{\tabcolsep}{-3pt}
    \resizebox{0.9\linewidth}{!}{
    \begin{tabular}{rccc|ccc|ccc}
    \toprule
        \multirow{2}{*}{Protocols} &\multicolumn{3}{c}{\codename-o3} & \multicolumn{3}{c}{\text{\codename-ds}}  &\multicolumn{3}{c}{\codename-hy}\\\cmidrule{2-10}
        & Field & Type & Fn. & Fields & Type & Fn. & Fields & Type & Fn. \\ \midrule
        Modbus & \textbf{0.99} & \textbf{0.81} & \textbf{1.00} & 0.98 & 0.74 & 0.98 & 0.93 & 0.61 & 0.81 \\ 
        S7comm & \textbf{0.81} & \textbf{0.80} & \textbf{0.64} & {0.64} & 0.39 & 0.26 & 0.76 & 0.49 & 0.31 \\ 
        Ethernet/IP & \textbf{0.80} & \textbf{0.69} & \textbf{1.00} & 0.42 & 0.64 & 0.97 & 0.46 & 0.65 & \textbf{1.00} \\ 
        DNP3 & \textbf{0.80} & \textbf{0.82} & \textbf{0.90} & 0.65 & 0.47 & 0.42 & 0.61 & 0.37 & 0.25  \\ 
        DNS & {0.72} & \textbf{0.71 }& \textbf{0.93} & \textbf{0.75} & 0.64 & 0.73 & 0.71 & 0.46 & 0.58 \\ 
        FTP & \textbf{0.66} & \textbf{0.72} & \textbf{0.77} & 0.63 & 0.53 & 0.66 & 0.64 & 0.32 & 0.59  \\ 
        TFTP & \textbf{1.00} & 0.35 & \textbf{1.00} & 0.71 & {0.53} & 0.95 & 0.68 & \textbf{0.56} & 0.81 \\ 
        HTTP & \textbf{0.60} & \textbf{0.62} & \textbf{0.70} & \textbf{0.60} & \textbf{0.62} & \textbf{0.70} & 0.11 & 0.01 & 0.03 \\ \midrule
        Average & \textbf{0.76} & \textbf{0.72} & \textbf{0.74} & 0.58 & 0.59 & 0.62 & 0.49 & 0.39 & 0.46 \\ \bottomrule
    \end{tabular}}
    \flushleft\scriptsize
    \codename-o3: the backbone of \codename is GPT-o3;\flushleft \codename-ds: the backbone of \codename is DeepSeek-v3.1-terminus \flushleft
    \codename-hy: the backbone of \codename is Hy3
\end{table}


Table~\ref{tab:undoc_perf} shows the results on undocumented protocols under different approaches.
\begin{table}[!ht]
    \centering
    \caption{Effectiveness on Undocumented Protocols.}\label{tab:undoc_perf}
    \addtolength{\tabcolsep}{-2pt}
    \resizebox{\linewidth}{!}{
    \begin{tabular}{rccc|ccc|ccc}
    \toprule
        \multirow{2}{*}{Protocols} &\multicolumn{3}{c}{\codename} &\multicolumn{3}{c}{BinPRE} & \multicolumn{3}{c}{Vanilla GPT-o3} \\\cmidrule{2-10}
        & Field & Type & Fn. & Fields & Type & Fn. & Fields & Type & Fn. \\ \midrule
        KTP & 0.50 & 0.22 & 0.44 & \textbf{0.44} & 0.00 & 0.00 & \textbf{0.57} & \textbf{0.24} & 0.13  \\
        CCN & \textbf{1.00} & \textbf{0.72} & \textbf{0.92} & 0.53 & 0.50 & 0.00 & 0.48 & 0.38 & 0.67 \\
        Xfer & \textbf{0.73} & \textbf{0.49} & \textbf{0.93} & 0.22 & 0.18 & 0.29 & 0.47  & 0.38 & 0.40\\ \midrule
        Avg. & \textbf{0.71}  & \textbf{0.43} & \textbf{0.77} & 0.41 & 0.19 & 0.11 & 0.52 & 0.31 & 0.43\\ \bottomrule
    \end{tabular}}
\end{table}

\subsubsection{Tables Related to RQ2}
Table~\ref{tab:fn_infer_results} show the details of metrics from \codename and baselines. DynPRE receives and responds to a subset of messages for S7comm and DNS, so the results depend on the messages it responds to.
\begin{table*}[htbp]\centering
\caption{Function Inference Results.}\label{tab:fn_infer_results}
\scriptsize
\resizebox{\linewidth}{!}{
\begin{tabular}{rlll|lll|lll|lll|lll|lllr}\toprule
\multirow{2}{*}{Protocols} &\multicolumn{3}{c}{SemPRE} &\multicolumn{3}{c}{BinPRE} &\multicolumn{3}{c}{Polyglot} &\multicolumn{3}{c}{Tupni} &\multicolumn{3}{c}{Binaryinferno} &\multicolumn{3}{c}{DynPRE} \\\cmidrule{2-19}
&Pre. &Rec. &F1 &Pre. &Rec. &F1 &Pre. &Rec. &F1 &Pre. &Rec. &F1 &Pre. &Rec. &F1 &Pre. &Rec. &F1 \\\midrule
Modbus &1.00 &1.00 &\textbf{1.00} &1.00 &1.00 &\textbf{1.00} &0.64 &1.00 &0.78 &1.00 &0.50 &0.67 &0.00 &0.00 &0.00 &0.00 &0.00 &0.00 \\
S7comm &0.78 &0.55 &\textbf{0.64} &0.00 &0.00 &0.00 &0.00 &0.00 &0.00 &0.00 &0.00 &0.00 &0.00 &0.00 &0.00 &$\text{0.00}^1$  &$\text{0.00}^1$  &$\text{0.00}^1$ \\
Ethernet/IP &1.00 &0.50 &\textbf{0.67} &1.00 &0.10 &0.17 &0.00 &0.00 &0.00 &0.00 &0.00 &0.00 &0.00 &0.00 &0.00 &N/A &N/A &N/A \\
DNP3 &0.99 &0.83 &\textbf{0.90} &1.00 &0.75 &0.86 &0.25 &0.25 &0.25 &1.00 &0.25 &0.40 &0.00 &0.00 &0.00 &N/A &N/A &N/A \\
DNS &1.00 &0.86 &\textbf{0.93} &0.80 &0.71 &0.75 &0.00 &0.00 &0.00 &0.00 &0.00 &0.00 &0.00 &0.00 &0.00 &$\text{0.00}^2$  &$\text{0.00}^2$  &$\text{0.00}^2$ \\
FTP &1.00 &0.63 &\textbf{0.77} &0.00 &0.00 &0.00 &0.00 &0.00 &0.00 &0.00 &0.00 &0.00 &0.00 &0.00 &0.00 &N/A &N/A &N/A \\
TFTP &1.00 &1.00 &\textbf{1.00} &1.00 &0.47 &0.64 &0.00 &0.00 &0.00 &0.00 &0.00 &0.00 &0.00 &0.00 &0.00 &N/A &N/A &N/A \\
HTTP &0.99 &0.54 &\textbf{0.70} &0.00 &0.00 &0.00 &$\text{0.00}^3$ &0.24 &$\text{0.00}^4$ &0.00 &0.00 &0.00 &0.00 &0.00 &0.00 &N/A &N/A &N/A \\
\midrule
Average &0.96 &0.61 &\textbf{0.74} &0.82 &0.22 &0.35 &0.01 &0.09 &0.01 &$\text{0.00}^5$ &0.04 &$\text{0.00}^6$ &0.00 &0.00 &0.00 &0.00 &0.00 &0.00 \\
\bottomrule
\end{tabular}}
\flushleft
Average - The performance on all protocols;\quad Bold numbers indicate the best results\\
1. counting based on the first 31 messages;\quad 2 - counting based on the first 16 messages;\quad 3 - 0.0015;\quad 4 - 0.0030;\quad 5 - 0.0026;\quad 6 - 0.0050
\end{table*}

\end{document}